\documentclass[manuscript]{aastex}
\usepackage{pdflscape}

\shorttitle{Properties of the open cluster Tombaugh~1}
\shortauthors{Sales Silva et al.}

\begin{document}


\title{Properties of the open cluster Tombaugh~1 \\
    from high resolution spectroscopy and uvbyCaH$\beta$ photometry\thanks{Based on observations carried out at
    Las Campanas Observatory (program ID: CN2009B-042) and Cerro Tololo Inter-American Observatory.}}


\author{Jo\~ao V. Sales Silva$^{1,2}$, Giovanni Carraro$^{1,3}$, Barbara J. Anthony-Twarog$^4$, Christian Moni Bidin$^5$, Edgardo Costa$^6$ and Bruce A. Twarog$^4$} 
\affil{1: ESO, Alonso de Cordova 3107, 19001, Santiago de Chile, Chile\\
       2: Observat\'orio Nacional/MCT, Rua Gen. Jos\'e Cristino, 77, 20921-400, Rio de Janeiro, Brazil\\
       3: On leave from Dipartimento di Fisica e Astronomia, Universit\'a di Padova, Italy.\\
       4: Department of Physics and Astronomy, University of Kansas, Lawrence, KS 66045-7582, USA\\
       5: Instituto de Astronomia, Universidad Cat\'olica del Norte, Av. Angamos 0610, Casilla 1280, Antofagasta, Chile\\
       6: Departamento de Astronomia, Universidad de Chile, Casilla 36-D, Santiago, Chile}
\email{joaovictor@on.br,gcarraro@eso.org,bjat,btwarog[@ku.edu],\\cmoni@ucn.cl,costa@das.uchile.cl}








\begin{abstract}
\par Open clusters can be the key to deepen our knowledge on various issues involving the structure and evolution of the Galactic disk and
details of stellar evolution because a cluster's properties are applicable to all its members. However the number of open clusters with 
detailed analysis from high resolution spectroscopy and/or precision photometry imposes severe limitation on studies of these objects.

\par To expand the number of open clusters with well-defined chemical abundances and fundamental parameters, we investigate 
the poorly studied, anticenter open cluster Tombaugh 1.

\par Using precision $uvbyCa$H$\beta$ photometry and high resolution spectroscopy, we derive the cluster's reddening, obtain 
photometric metallicity estimates and, for the first time, present detailed abundance analysis of 10 potential 
cluster stars (9 clump stars and 1 Cepheid).

\par Using radial position from the cluster center and multiple color indices, we have isolated a sample of unevolved probable, 
single-star members of Tombaugh 1. From 51 stars, the cluster reddening is found to be $E(b-y)$ = 0.221 $\pm$ 0.006 or $E(B-V)$ =
0.303 $\pm$ 0.008, where the errors refer to the internal standard errors of the mean. The weighted photometric metallicity from
$m_1$ and $hk$ is [Fe/H] = $-$0.10 $\pm$ 0.02, while a match to the Victoria-Regina Str\"{o}mgren isochrones leads to an 
age of 0.95 $\pm$ 0.10 Gyr and an apparent modulus of $(m-M)$ = 13.10 $\pm$ 0.10. Radial velocities identify 6 giants 
as probable cluster members and the elemental abundances of Fe, Na, Mg, Al, Si, Ca, Ti, Cr, Ni, Y, Ba, Ce, and Nd have been derived
for both the cluster and the field stars.

\par Tombaugh 1 appears to be a typical inner thin disk, intermediate-age open cluster of slightly subsolar metallicity, located just
beyond the solar circle, with solar elemental abundance ratios except for the heavy s-process elements, which are a factor of 
two above solar. Its metallicity is consistent with a steep metallicity gradient in the galactocentric region between 9.5 and
12 kpc. Our study also shows that Cepheid XZ CMa is not a member of Tombaugh 1, and reveals that this Cepheid presents signs of barium enrichment, making it a probable
 binary star.
\end{abstract}





\section{Introduction}

Tombaugh~1 is a star cluster in the Canis Majoris constellation at  $\alpha= 07^{h}00^{m}29^{s}$, 
$\delta= -20^{o}34^{\prime}00^{\prime\prime}$, and $\ell,b=232^{\circ}.22,-7^{\circ}.32$ (2000.0 equinox), 
discovered in 1938 by Clyde Tombaugh (Tombaugh 1938). A quick glance at any sky 
image shows only a small enhancement of stars in a rich Galactic field, a primary reason why Tombaugh~1 has been 
studied very little until now. As detailed in Sect. 2, estimates of its fundamental parameters vary strongly 
from author to author, photometric data are scanty, and spectroscopic data totally absent.

There are, however, several reasons to consider this cluster particularly interesting. With a Galactic latitude of $-7^{o}.32$ and an 
assumed a distance of 2--3 kpc (Piatti et al. 2004; Carraro \& Patat 1995), Tombaugh 1 is situated $\sim$300--400 pc below 
the Galactic plane, a location rare among open clusters of the presumed age of Tombaugh~1 ($\lesssim 10^9$ yr).  
The direction to Tombaugh~1 intersects the Perseus and Orion arms in the third Galactic quadrant, and the cluster 
was for some time associated with the putative Canis Major dwarf galaxy (Bellazzini et al. 2004). Although the 
discussion of this topic has been dormant for some time (Carraro et al. 2008), it would still 
be valuable to determine the cluster distance with enough precision to associate it with either the Perseus or, 
possibly, the Orion arm extension in the third Galactic quadrant (V\'asquez et al. 2008). Moreover, a distance of 2--3 kpc places 
Tombaugh~1 at a galactocentric distance of 10--11 kpc, a location where some observations indicate an abrupt 
change in the abundance gradient of the disk (see, e.g. Twarog et al. 1997; Lepine et al. 2011). 
Improved cluster distance and metallicity estimates would be of paramount importance in probing this picture of the outer 
Galactic disk chemical properties. Metallicity, in particular, has never been measured reliably, but only inferred 
either from photometric indices or via comparison with isochrones (Piatti et al. 2004).

With these goals in mind, we present new Str\"omgren $uvbyCa$H$\beta$ photometry for a large field around the cluster 
and high resolution spectroscopy of 10 potential cluster stars (9 red giant clump stars and 1 Cepheid). This unique dataset will be used to provide precise 
estimates of the cluster basic parameters, in particular, reddening, distance, age, overall and elemental metallicity.

The paper is organized as follows: Section~2 gives details on the limited investigation of the cluster to date. In Section~3 
we present the photometric and spectroscopic observations and their reduction, as well as a description of the radial-velocity
determinations for membership. Section 4 details the derivation of the cluster properties from photometry, while Section~5 is devoted 
to the spectroscopic abundance analysis. In Section~6 we discuss the results of our spectroscopic analysis in detail.
Section~7 interprets Tombaugh 1 in the context of galactic evolution and summarizes our findings.

\section{Tombaugh~1: Background}
Tombaugh discovered Tombaugh~1 and Tombaugh~2 during the trans-Neptunian planet search at Lowell Observatory (Tombaugh 1938, 1941),
reporting an apparent diameter for Tombaugh 1 of $\sim5^{\prime}$, with typical cluster stars having  a visual magnitude of 14-15. 
His short description of the cluster notes that the field in the direction of Tombaugh~1 is extremely rich, suggesting
that this might be the reason why the cluster went undetected for so long. 

As discussed in Turner (1983); Haffner (1957) and Tifft (1959) independently rediscovered Tombaugh~1, reporting 
discrepant values for the cluster declination; the Haffner (1957) position is incorrect. Tifft (1959) noted the 
cluster because the Cepheid XZ CMa lies about one cluster diameter northward of the cluster center.
Turner (1983) provided the first estimates for the cluster fundamental parameters from analysis of $UBV$ photoelectric 
photometry of 26 stars, including 5 likely members and 10 possible members selected by radial location in the cluster
and position in the two-color diagram, the $UBV$ diagram also indicating $E(B-V)$ = 0.27 $\pm$ 0.01 mag. Turner (1983) measured a 
cluster diameter of $\sim$10$^{\prime}$ and estimated a distance and age of 1.26 kpc ($(m-M)$ = 11.34 $\pm$ 0.04) 
and $\sim$800 Myr, respectively, using the Hyades cluster adjusted to solar metallicity as a reference, 
though the sparse photometry extended barely 1.5 mag below the top of the turnoff. Lastly, Turner (1983) suggested 
that Tombaugh~1 hosts a probable blue straggler star (BSS), later confirmed by Ahumada \& Lapasset (1995).

Turner (1983) also investigated the membership of Cepheid XZ CMa with Tombaugh 1. Turner suggested that XZ CMa is not a member of Tombaugh 1. However, this conclusion was never subjected to a more rigorous
analysis based in high-resolution spectroscopy. So, we also analyzed spectroscopically XZ CMa to confirm or exclude the
cluster membership. In fact, cluster Cepheids are important to fix the distance scale (see e.g., Majaess et al. 2013a;
Majaess et al. 2013b).

The first CCD study of Tombaugh 1, limited to $VI$, was carried out by Carraro \& Patat (1995), covering an area 6$^{\prime}$ on a side, essentially 
the cluster core (Turner 1983). Very different values for some of the cluster basic parameters were found: reddening, distance, 
and age of $E(B-V)$ = 0.40 $\pm$ 0.05, 3 kpc ($(m-M)$ = 13.60 $\pm$ 0.2), and 1 Gyr, respectively, tied to color-magnitude diagram (CMD)
matches to theoretical isochrones. While this study dealt with only two filters and field star contamination makes it difficult to 
identify the cluster turnoff clearly, the parameter differences are not unexpected given the sparse sample of the earlier study.

Piatti et al. (2004) presented a more extensive CCD study using Washington photometry, covering a large area around the cluster.  
The cluster was found to be 1.3 Gyr old from a combination of CMD morphology and isochrone fits, assuming [Fe/H] = $-$0.40, 
with distance and reddening estimates intermediate between the Turner (1983) and Carraro \& Patat (1995) values. An attempt 
was also made to directly measure the metallicity using Washington photometry, obtaining [Fe/H] = $-$0.30
with a large uncertainty of $\pm$0.25 dex.

Finally, on the basis of stellar data from PPMXL\footnote{PPMXL is a catalog of positions, proper motions, 
2MASS, and optical photometry of 900 million stars and galaxies. For more information: \url{http://vo.uni-hd.de/ppmxl}} 
and 2MASS, Kharchenko et al. (2013) obtained some spatial, structural, kinematic, and astrophysical parameters of Tombaugh 1. 
In particular, they determined for Tombaugh 1 an age of 1.16 Gyr, a reddening $E(B-V)$ = 0.281 and a distance of 2642 pc
($(m-M)$ = 12.98), values similar to those obtained by Piatti et al. (2004). Kharchenko et al. (2013) also estimated 
average proper motion in right ascension ($-$0.99 mas/yr) and in declination (3.97 mas/yr), but didn't estimate the 
average radial and galactic space velocities of Tombaugh 1.

\section{Observations and Data Reduction}
\subsection{Photometry}
\noindent
Photometry for Tombaugh~1 was secured in December 2010, during a 5-night run using the Cerro Tololo Inter-American Observatory 
1.0m telescope operated by the SMARTS consortium\footnote{\tt http://http://www.astro.yale.edu/smarts}. The telescope is 
equipped with an STA~$4064\times4064$ CCD camera\footnote{\texttt{http://www.astronomy.ohio-state.edu/Y4KCam/detector}}
with 15-$\mu$m pixels, yielding a scale of 0.289$^{\prime\prime}$/pixel and a field-of-view (FOV) of $20^{\prime} \times 20^{\prime}$ 
at the Cassegrain focus of the telescope.

In Table~1 we present the log of our  Str\"omgren observations, together with exposure times and airmasses. A total of 75 images
were acquired for Tombaugh~1. All observations were carried out under photometric conditions with good-seeing (0.8--1.2 arc sec). 
A sample image of the covered field is shown in Fig.~1.

Basic calibration of the CCD frames was done using the Yale/SMARTS y4k reduction script based on the 
IRAF\footnote{IRAF is distributed by the National Optical Astronomy Observatory, operated by the Association of 
Universities for Research in Astronomy, Inc., under cooperative agreement with
the National Science Foundation.} package \textsc{ccdred}, and the photometry was performed
using IRAF's \textsc{daophot} and \textsc{photcal} packages. Instrumental magnitudes were extracted following
the point spread function (PSF) method (Stetson 1987) using a quadratic, spatially-variable
master PSF (PENNY function). Finally, the PSF photometry was aperture-corrected using 
corrections determined from aperture photometry of bright, isolated stars in the field.

Standard stars for the extended Str\"omgren system were observed on one of the photometric nights on which Tombaugh 1 was observed. 
We additionally employed observations of secondary standard fields in several open clusters, using the same telescope and instrument 
one year later, to derive the form of the calibration equations. The clusters observed in December 2011 were M67 
(Nissen et al. 1987), NGC 2287 (Schmidt 1984) and NGC 2516 (Snowden 1975).   
The zeropoint for each calibration equation applied to Tombaugh 1 was anchored by observations of 
eight field star standards obtained on 7 December, 2010. Standard values were obtained from the catalogs 
of Olsen (1983, 1993, 1994) for $uvby$, from Hauck \& Mermilliod (1998) for H$\beta$ values and from 
Twarog \& Anthony-Twarog (1995) for $hk$ index values for the field star standards. 

Table 2 summarizes the calibration equations' slopes and color terms.  Following a standard practice for Str\"omgren photometry, 
calibration of $(b-y)$ would require that separate slopes be determined for cooler dwarfs as distinct from warmer dwarfs and 
giants. Insufficient cool dwarf standards were observed to accomplish this.  The slope listed 
in Table 2 is appropriate for giants and dwarfs with $(b-y)_0 \leq 0.42$; application of this 
slope to dwarfs redder than this is an unavoidable extrapolation.  Bluer dwarfs represent 
the only class for which $m_1$ and $c_1$ calibrations could be established with any confidence. 
The errors listed in Table 2 represent the standard deviation of the calibrated values 
about the standard values for the field star standards, indicating the external precision 
of the zeropoints of the calibration equations. The final calibrated catalog was then 
cross-correlated with 2MASS to convert pixel (i.e., detector) coordinates into RA and DEC 
for J2000.0 equinox, thus providing 2MASS-based astrometry. An excerpt of the optical 
photometric table used in this investigation is illustrated in Table 3.
Fig. 2 shows the trend of errors with $V$ magnitude. The $V$ and $b-y$ data remain below 0.02 mag uncertainty to the
limit of Table 3 ($V$ = 18.5), while errors in the remaining indices begin to rise above this value at $V$ = 16.5, 17.25,
17.25, and 17.5 for $c_1$, $m_1$, $hk$, and H$\beta$, respectively.

\subsection{Spectroscopy and Radial Velocities}

Over the night of January 5, 2010, we observed ten potential 
cluster stars (nine clump stars and one Cepheid, see Sec. 4.1)  
with the \textit{Inamori-Magellan Areal Camera \& 
Spectrograph} (IMACS, Dressler et al. 2006) attached to the Magellan telescope (6.5 m) located at Las Campanas, Chile. The spectra were 
obtained using Multi-Object Echelle (MOE) mode with two exposures, one of 900s and other of 1200s. Our spectra have a resolution 
of R$\sim$20000, while the spectral coverage depends on the location of the star on the multi-slit mask, but generally goes
from 4200 \AA{} to 9100 \AA{}. The detector consists of a mosaic with eight CCDs with gaps of about 0.93 mm between the CCDs, causing 
small gaps in stellar spectra. 

The spectra were reduced following the standard procedures using IRAF, which includes CCD 
bias and flat-fielding correction, spectrum extraction, wavelength calibration and sky subtraction using the tasks \textsc{ccdproc}, 
\textsc{doecslit}, \textsc{ecidentify} and \textsc{background}, respectively. For each CCD, we performed bias and flat-fielding correction 
separately, after which we used the IRAF tasks \textsc{imcreate} and \textsc{imcopy} to join the CCDs and create the mosaic. The cosmic rays 
were removed with the IRAF Laplacian edge-detection routine (van Dokkum 2001), and the radial velocities were obtained from the wavelength
shift of the unblended absorption lines of Fe covering the
entire wavelength range. The values of wavelength shift were measured via line-by-line comparison between observed and
laboratory wavelength with the lines center of observed wavelength being determined through the task \textsc{splot} in
IRAF. To derived the final radial velocities we applied a zero-point offset correction using the task \textsc{fxcor} in
IRAF to cross-correlate the telluric lines of the observed spectra with telluric lines of the high-resolution FEROS solar
spectrum collected by us in a previous run (Moni Bidin et al. 2012). To calculate the heliocentric velocities and combine
the spectra of different exposures, we used the IRAF tasks \textsc{rvcorrect} and \textsc{scombine}, respectively. We took the star’s 
heliocentric radial velocity to be the average of the two epochs measured and the error to be the difference between 
the two values multiplied by 0.63 (small sample statistics; see Keeping 1962). The nominal S/N ratio was evaluated by
measuring with IRAF the rms flux fluctuation in selected continuum windows. The values at 6000 \AA{} 
are shown in Table \ref{sample}.

Table \ref{sample} gives some information about the observed stars: IDs (Carraro \& Patat 1995), right ascension, declination, 
$V$ and $b-y$ from Table 3 and $V-I$ photometry (Carraro \& Patat 1995), heliocentric radial velocities (RV$_{1}$ and RV$_{2}$) 
at two epochs and  their mean values 
($\langle$RV$\rangle$), projected rotational velocities (vsini) and spectral signal-to-noise at 6000 \AA{}. We estimated 
the projected rotational velocities, vsini, by a spectral synthesis technique using unblended Fe lines analyzed with  
model atmospheres, a macroturbulent  velocity of 3 km s$^{-1}$, limb darkening and instrumental 
broadening corresponding to IMACS spectral resolution. For some stars with low vsini it was possible to determine 
only an upper limit because of the insensitivity of the spectral synthesis to vsini below 2.7 km s$^{-1}$ . 

In the literature, there is no information about the radial velocity of Tombaugh 1. So, to determine the membership of
stars, we first found for a group of stars with similar heliocentric radial velocities (RV) in the sample, to have a
preliminary cluster radial velocity and a list of members, that could be iteratively refined. The stars with RV within
2$\sigma$ of the cluster mean heliocentric radial velocity were classified as member of Tombaugh 1. The membership of the
Cepheid XZ CMa (star 806) was not defined by its heliocentric radial velocity, because its RV is variable due to pulsations. 
So, we classified XZ CMa as non-member of Tombaugh 1 because its metallicity ([Fe/H]=$-$0.53) is much lower than
metallicity of stars classified as member of Tombaugh 1 (see Table 9). We identify six red clump 
giants belonging to Tombaugh 1 and derive a mean cluster heliocentric radial velocity of 81.1 $\pm$ 2.5 km s$^{-1}$.

\section{Cluster Parameters from Photometry}
\noindent
As discussed previously, one of the primary challenges in identifying and studying Tombaugh 1 is isolating the moderately populated 
cluster from the rich stellar background. This is particularly important for defining directly the fundamental cluster parameters of
reddening and metallicity and indirectly the distance and age. The challenge is illustrated in Fig. 3, where the $V, b-y$ CMD for the
entire field of study is presented. Red giants observed as part of this investigation and found to
be probable radial-velocity members are plotted as red starred points; probable nonmember are open red triangles. 
The complexity of the 
stellar population mix in this region of the galaxy is evident and will be discussed in detail in a future paper.
To enhance our definition of the cluster, we first reinvestigate the radial profile of the cluster.

\subsection{Star Counts and Cluster Size}
\noindent
To quantify the amount of field star contamination, we performed star counts to derive an estimate of 
the cluster center and size. Using an array of positions covering the field of the CCD, we derived 
a density contour map and calculated the density inside each grid step by a kernel estimate (Carraro et al. 2014c).
This is shown in Fig.~4, which confirms the appearance of Fig.~1 that Tombaugh~1 is far from 
being a symmetric object. The cluster looks elongated in the direction NE to SW, and the highest peak does
not represent the center of a uniform star distribution. The largest peak is located at RA = 105$^{o}$.11, 
DEC = $-$20$^{o}$.58, while the nominal center of the cluster is clearly displaced to the northeast direction 
at RA = 105$^{o}$.13, DEC = $-$20$^{o}$.54. The loose and irregular shape of Tombaugh~1 may be the result 
of its dynamical evolution due to its tidal interaction with the Milky Way. However, little kinematic 
information beyond the cluster radial velocity exists to confirm this scenario. High quality proper motions 
could go a long way to defining the direction of the cluster motion and test if this
coincides with the direction of the apparent cluster deformation, indicating if Tombaugh~1 has 
indeed been tidally disturbed.

To isolate probable cluster members, assumed to be those stars which lie within the cluster boundaries, 
we derive the cluster radial surface density profile shown in Fig.~5. This has been computed by 
drawing concentric rings centered on the nominal cluster center. This is motivated by the fact that, 
while the densest central regions look distorted, the cluster halo still retains a more spherical profile.
Star counts level off at $\sim$4$^{\prime}$ from the cluster nominal center, close to the value reported 
by Turner (1983). The mean density in the field surrounding the cluster is 5 stars/arcmin$^2$ (see also Fig.~4), 
and our survey covers the whole cluster area. As a consequence, in the following we will adopt 4$^{\prime}$ as the 
cluster radius and refer to this area as the cluster area, while the area outside 4$^{\prime}$ from the cluster center 
will be referred to as the offset field.

\subsection{Photometric Reddening and Metallicity}
\noindent
In the absence of membership information for any stars beyond those in Table 4, we can enhance the likelihood of including
cluster members in our sample by restricting the analysis to stars within 3.5$\arcmin$ of the cluster center, just short of
the transition region from the cluster to the field based upon star counts. Fig. 6 shows the $V, b-y$ CMD for stars within this core;
all red giant members, independent of radial location, are plotted as stars, while open triangles are probable
nonmembers. The cluster's turnoff region and the blue edge of the main sequence are well-defined to the limit of the survey. 
The color spread at the top of the turnoff and the color and magnitude differential between the turnoff and the giant branch 
are very reminiscent of NGC 5822, a cluster of slightly subsolar metallicity with an age of 0.9 Gyr (Carraro et al. 2011).
An additional means of demonstrating the cluster population comes from the $V, hk$ CMD for the core region, shown in Fig. 7.
In the likely probability that there is a modest reddening range across the face of the cluster, this CMD minimizes the impact
due to the weak sensitivity of $hk$ to reddening but a strong sensitivity to temperature and metallicity changes, factors we
will make effective use of below. The tight vertical band defining the cluster turnoff reflects this fact, while the 
steep slope in $V$ with $hk$ is indicative of the cluster age; the trend in $hk$ with decreasing $b-y$ plateaus as the 
stellar sample moves from F to A stars, leading to an almost vertical turnoff in the $V, hk$ CMD for clusters of 
intermediate age, as illustrated by NGC 5822 in Fig. 13 of Carraro et al. (2011). Among the giants, the positions of the two
faintest radial-velocity members, 663 and 1349, place them redward of the majority of the probable clump stars and
indicates that, despite their similar velocities, they are probable field stars. In the absence of more information, they will
be retained in the discussions below.

To further isolate probable members for defining the reddening and metallicity, we restrict our sample to
stars populating the blue edge of the cluster main sequence between $V$ = 15.50 and 17.0. The bright cutoff eliminates 
the evolved stars that populate the color spread at the top of the main sequence while the faint boundary defines the
magnitude range where errors in the color indices begin to increase for $m_1$, $hk$, and especially $c_1$. The blue edge 
of the main sequence in this magnitude range was used to define a single-star boundary and any star within $\sim$0.4 mag of the
boundary was classified as a single star (blue circles in Fig. 8). Stars between 0.4 and 0.8 mag were classed as
probable binaries (black squares in Fig. 8), if members, while all stars more than 0.8 mag beyond the main sequence were 
tagged as likely field stars (red crosses in Fig. 8). 

We can check this classification using the $V, hk$ CMD of Fig. 9. Stars can be located off the main sequence
for a variety of reasons: bad and/or contaminated photometry, binarity, excessive reddening compared to
the typical cluster star and/or nonmembership. As noted earlier, the $hk$ index is very sensitive to color changes due
to temperature and only weakly impacted by reddening. In fact, increased reddening moves a star blueward in $hk$. 
As shown in Fig. 9, the separation by class as defined by Fig. 8 is well corroborated. With only four obvious
exceptions, the single stars form a well-defined turnoff band covering a modest range in $hk$. The stars 
redward of the main sequence band in Fig. 9 are dominated by the
stars tagged as redder in Fig. 8, indicating that these are truly redder than the cluster sequence and that the majority are probable
nonmembers. Three red crosses which sit on the main sequence in Fig. 9 deserve some explanation. These are either highly reddened
field stars or, more likely, field stars in the direction of the cluster with significantly lower metallicity than Tombaugh 1. 
For metallicity and reddening estimation, we will limit the sample to the single stars (blue circles), with the four points which deviate
from the cluster main sequence in Fig. 9 excluded.

For consistency with past cluster work, we will derive the reddening from two Str\"omgren relations from Olsen (1988) and Nissen (1988), 
a slightly modified version of the original relations derived by Crawford (1975, 1979). Reddening estimates are derived 
in an iterative fashion. The indices are corrected using an initial guess at
the cluster reddening and the intrinsic $b-y$ is derived using the reddening-free H$\beta$ adjusted for metallicity and evolutionary
state. A new reddening is derived by comparing the observed and intrinsic colors and the procedure repeated. The reddening
estimate invariably converges after 2--3 iterations. To derive the reddening, one needs to correct $b-y$ for metallicity, so
a fixed [Fe/H] is adopted for the cluster and the reddening derived under a range of [Fe/H] assumption that bracket the final
value. The complementary procedure is to vary the mean reddening value for the cluster and derive the mean [Fe/H]. Ultimately,
only one combination of $E(b-y)$ and [Fe/H] will be consistent.

For Tombaugh 1, the metallicity from $m_1$ was varied between [Fe/H] = $-$0.28 and $+$0.12, generating a range of 
$E(b-y)$ = 0.224 to 0.214 for the relation of Olsen (1988) and 0.223 to 0.216 for Nissen (1988) from 51 stars
within the H$\beta$ calibration range. In all cases, the standard error of the mean for the final $E(b-y)$ is $\pm$ 0.006 mag. 
For [Fe/H] from $m_1$ equal to $-$0.16, the reddening from the two relations is virtually identical at 
$E(b-y)$ = 0.221 $\pm$ 0.006; the difference between the two reddening values is statistically insignificant 
compared to the standard errors of the mean.  If $E(b-y)$ = 0.73*$E(B-V)$, the reddening estimate from Str\"omgren data 
alone is $E(B-V) = 0.303 \pm 0.008$. There is weak evidence for a variation in $E(b-y)$ across the face of the cluster,
with the reddening being higher on average by 0.03 mag in the southwest and lower by a comparable amount in the northeast.
Without more membership information, for purposes of deriving the cluster parameters, we will adopt the cluster mean for
all stars.

With the reddening set, we can derive the metallicity from both $m_1$ and $hk$, using H$\beta$ as the primary temperature index.
From 51 stars, [Fe/H] = $-$0.165 $\pm$ 0.027 from $m_1$ and $-$0.086 $\pm$ 0.013 from $hk$. If one anomalous
measurement located more than three sigma from the cluster mean is removed from the $m_1$ analysis, the revised [Fe/H] becomes
$-$0.153 $\pm$ 0.025. The greater uncertainty in the metallicity estimate from $m_1$ relative to $hk$ is a reflection of the
greater sensitivity of $m_1$ to reddening changes and a lower sensitivity to metallicity variation; the small difference in
[Fe/H] can be entirely explained by a zero-point offset of 0.005 mag in $m_1$. Weighting the two photometric estimates by the
inverse square of the errors leads to a final value of [Fe/H] = $-$0.10 $\pm$ 0.02.

\subsection{Age and Distance Estimation}
\noindent
One of the rare sets of available isochrones which include models transformed to the Str\"omgren system is the 
Victoria-Regina (VR) set of isochrones (VandenBerg et al. 2006). Fig. 10 shows the scaled-solar models 
for [Fe/H] = $-$0.11, ages 0.8, 0.9, and 1.0 Gyr, adjusted for $E(b-y)$ = 0.221 and $(m-M)$ = 13.10. 
Symbols have the same meaning as in Fig. 6. The already noted similarity of Tombaugh 1 to NGC 5822 is confirmed.
In addition to the similar scatter in $b-y$ at the top of the turnoff, the best fit age estimate is between 0.9 and 1.0 Gyr;
the best fit to a different set of broad-band isochones for NGC 5822 produced an age of 0.9 $\pm$ 0.1 Gyr for
the more populated cluster (Carraro et al. 2011). The reddening-corrected true distance modulus is $(m-M)_o$ = 12.15, 
in excellent agreement with the most recent work of Kharchenko et al. (2013).

\section{Atmospheric Parameters and Abundances Analysis}
\noindent
The equivalent width measurements of absorption lines of Na, Mg, Al, Si, Ca, Ti, Cr, Ni and Fe were used to obtain their 
respective chemical abundances while the abundances of Y, Ba, Ce and Nd were derived through spectral synthesis. The 
equivalent widths were measured manually using the task \textsc{splot} in IRAF to fit a Gaussian profile to the observed
absorption line. We rejected the absorption lines with equivalent widths greater than 160 m\AA{} because these lines 
are saturated, which prevents a Gaussian fit to the absorption lines (Pereira et al. 2011). All equivalent widths used 
to obtain the atmospheric parameters and chemical abundances are shown in the Tables \ref{tabelFea}, \ref{tabelFeb}, 
\ref{tabellinesa} and \ref{tabellinesb}.

The atomic-line list adopted in this work is the same as the one used by Santrich et al. (2013) and 
Sales Silva et al. (2014). For Ba II line, the hyperfine structure (HFS) was taken into account and we
used the line list of Carraro et al. (2014b). In Tables \ref{tabelFea} 
and \ref{tabelFeb} we describe the line list with excitation potential ($\xi$) and oscillator strength ($gf$) for absorption 
lines of Fe\,{\sc i} and Fe\,{\sc ii}. The values ​​of the oscillator strength adopted for the Fe\,{\sc i} and Fe\,{\sc ii} lines were taken from 
Lambert et al. (1996) and Castro et al. (1997). Tables \ref{tabellinesa} and \ref{tabellinesb} show the atomic parameters
($gf$ and $\xi$ values) of the absorption lines of the elements Na, Mg, Al, Si, Ca, Ti, Cr and Ni with their 
respective references (column 5). Atomic parameters for several transitions of Ti, Cr, and Ni were retrieved from 
the National Institute of Science and Technology Atomic Spectra Database (Martin et al. 2002). For Na we used only two absorption
lines, 6154.226 \AA{} and 6160.747 \AA{}. These Na lines have a clean profile which makes it possible to calculate 
the chemical abundance of Na through the equivalent width (Smiljanic 2012). The absorption lines used 
to obtain s-process elements abundances were 5289 \AA{} and 5402 \AA{} for Y, 5853 \AA{} for Ba, 5117 \AA{} and
5187 \AA{} for Ce and 4914 \AA{} for Nd.

The atmospheric parameters and chemical abundances were obtained in the same manner as in Pereira et al. (2011),
Santrich et al. (2013), and Sales Silva et al. (2014) using the local thermodynamic equilibrium (LTE) model 
atmospheres of Kurucz (1993) and the spectral analysis code MOOG (Sneden 1973). 
Excitation equilibrium was used to derive the effective temperature ($T_{\rm eff}$) as defined by a zero slope 
of the trend between the iron abundance derived from Fe\,{\sc i} lines and the excitation potential of the measured lines. 
Microturbulent velocity was adjusted until both the strong and weak Fe\,{\sc i} lines (represented by reduced equivalent width, 
$W_{\lambda}/{\lambda}$) gave the same abundance. Finally, the surface gravity was determined using the 
ionization equilibrium found from the equality of the abundances of
Fe\,{\sc i} and Fe\,{\sc ii}. The final adopted atmospheric parameters are given in Table \ref{tab:atmparam}.

The uncertainty in the slopes of the Fe\,{\sc i} abundance versus excitation potential and Fe\,{\sc i} abundance versus 
reduced equivalent width were used to derive the uncertainties in our adopted effective temperatures ($T_{\rm eff}$) and 
microturbulent velocities ($\xi$), respectively. The standard deviation in $\log g$ was set by changing
this parameter around the adopted solution until the difference between Fe\,{\sc i} and Fe\,{\sc ii} mean 
abundance differed by exactly one standard deviation of the [Fe\,{\sc i}/H] mean value. We estimated typical uncertainties 
in atmospheric parameters of the order of $\pm$180\,K, $\pm$0.3~dex, and $\pm$0.3 km\,s$^{-1}$ for
$T_{\rm eff}$, $\log g$ and $\xi$, respectively.

We also calculated the photometric effective temperature and photometric gravity to compare with our spectroscopic temperature 
and gravity. Photometric temperatures were calculated using the calibration of Alonso et al. (1999) and our values of 
$(b-y)$ with $E(b-y)$ = 0.221. The photometric gravity for each star was obtained from the equation:

\begin{eqnarray}
\log g_{\star}\;  & = & \log \frac{M_{\star}}{M_{\odot}}  
+ 0.4\left(V-A_{ V}+BC\right)  \nonumber \\
& &  {{\,}\atop{\,}} + 4\log T_{\rm eff} - 2 \log r\; ({\rm pc}) - 10.62.
\end{eqnarray}

\noindent
Where $T_{eff}$ is the photometric effective temperature and $M$ is the mass.
Based upon the VR isochrones and an age of 0.95 Gyr, the typical mass for a star in the color range of the likely member red
giants is $2.15 M_\odot$. The photometric data of Table 4 were combined with an adopted distance of $r = 2700$ pc and 
bolometric corrections ($BC$) defined by the relations of Alonso et al. (1999). For the Sun we 
adopted $M_{bol \odot} = 4.74$ mag (Bessell et al. 1998), $T_{\rm eff \odot} = 5700$ K and $\log g_\odot = 4.3$ dex.

It should be emphasized that for the nonmembers stars, adoption of the cluster parameters for distance, reddening, and
metallicity will likely generate discordant results when compared to the spectroscopic parameters. For the six probable members,
the temperature difference, in the sense (spectroscopic - photometric), is 52 $\pm$ 196 K, while the residuals in $\log g$ are 0.22 $\pm$ 0.33,
consistent with the probable uncertainties in the estimates from the spectra, discussed above, and from the photometry. 
The modest offsets in temperature and gravity between the spectroscopic and photometric approaches are typical of such
comparisons. Different methods are known to produce systematic offsets from each other, but there is no consensus on the
source these offsets (e.g. Allende Prieto et al. 1999; Frebel et al. 2013). 

The determination of the atmospheric parameters (Table \ref{tab:atmparam}) and the knowledge of the 
atomic parameters of the absorption lines enables us to obtain the chemical abundance by measuring 
the equivalent widths or by spectral synthesis. In the case of equivalent widths, MOOG uses atmospheric 
and atomic parameters, as well as equivalent widths measurements, to calculate the chemical abundance. For 
spectral synthesis, as input for MOOG we supply the atmospheric and atomic parameters and an estimate of the 
chemical abundance of the elements that influence the absorption line studied. Thereafter MOOG 
generates a synthetic spectrum which is compared with the observed spectrum, iterating until
we find a chemical abundance that makes the synthetic spectra and observed identical. 

Tables \ref{abunda-Na} and \ref{abunda-Ni} show the chemical abundances of Na, Al, Fe-peak, 
alpha and s-process elements in the notation [X/Fe] and its standard deviation. We
analyzed a high-resolution FEROS solar spectrum to obtain the atmospheric parameters and 
solar abundance with the same methodology applied to red clump stars of Tombaugh 1. We found the 
following values for solar atmospheric parameters: $T_{\rm eff \odot}$ = 5700 K, $\log g_\odot$ = 4.3 dex 
and $\xi_\odot$ = 0.9 km\,s$^{-1}$. Pavlenko et al. (2012) found similar values of 5777 K, 
4.44 dex and 0.8 km\,s$^{-1}$ for the effective temperature, surface gravity and
microturbulent velocity, respectively. In Table \ref{sun} we show our solar chemical 
abundances together with those given by Grevesse \& Sauval (1998) and Asplund et al. (2009) 
for comparison. The adopted abundances for the elements analyzed in this work were normalized to our solar
abundances. In the seventh row of Tables \ref{abunda-Na} and \ref{abunda-Ni} we show the 
mean chemical abundance of Tombaugh 1 for each element with their respective standard deviations.

The approach to estimate the uncertainties in abundance consists in determining how the abundances for 
each element react to the errors associated with each atmospheric parameter, independent of the others. After 
that we combine quadratically all these errors and set this result as the total abundance uncertainty.
These total uncertainties are given in the 5th column of Table \ref{error} for star 769. We chose the star 769
to determine the abundance uncertainties for being one of the cluster giants that had the greatest number 
of elements with derived chemical abundance. The uncertainties for the aluminium weren't obtained for star 
769 because their absorption lines are located in the spectral gaps, so we used star 663 to calculate the 
aluminium uncertainties.  The uncertainties in abundance for the other stars generate similar values.

\section{Results of Abundance Analysis}

In this section we discuss the results of our chemical analysis via comparison with the chemical abundances of field giants 
stars and open clusters from the literature.

\subsection{Metallicty and Iron-peak Elements}

In Table \ref{tab:atmparam} we show the metallicities obtained for our giants. The range of metallicity for the six stars 
classified as members of Tombaugh 1 is $-$0.16 to 0.10 dex, with the mean of $-$0.02 $\pm$ 0.05 dex. The spectroscopic values 
are consistent with photometric value of $-$0.10 $\pm$ 0.02. 
A weighted average of the two approaches gives a final [Fe/H] = $-$0.08 for Tombaugh 1. Comparison with past abundance estimates
provides little insight given the large uncertainty in previous published estimates of this cluster parameter. 

In Figures 11, 12 and 13 we show the abundance ratio of [X/Fe] versus metallicity for our sample of giants, 
for giants from Mishenina et al. (2006) and Luck \& Heiter (2007), and also for the open clusters: NGC 6192, NGC 6404 and 
NGC 6583 (Magrini et al. 2010); NGC 3114 (Santrich et al. 2013); NGC 2527, NGC 2682, NGC 2482, NGC 2539, NGC 2335, 
NGC 2251 and NGC 2266 (Reddy et al. 2013); Trumpler 20 (Carraro et al. 2014b);
NGC 4337 (Carraro et al. 2014d); NGC 4815 and NGC 6705 (Magrini et al. 2014); 
Cr 110, Cr 261, NGC 2477, NGC 2506 and NGC 5822 (Mishenina et al. 2015).
For the s-process elements of Fig. 13, we added data of the open clusters Berkeley 25, Berkeley 73, Berkeley 75, 
Ruprecht 4, Ruprecht 7, NGC 6192, NGC 6404 and NGC 6583 from Mishenina et al. (2013).

From our chemical analysis of Tombaugh 1 we derive the following mean abundance ratios [X/Fe] for Cr 
and Ni: 0.10$\pm$0.06 and $-$0.04$\pm$0.02 dex, respectively. Our [Cr/Fe] and [Ni/Fe] of Tombaugh 1 are in 
good agreement with disk field giants and open clusters from literature as demonstrated in Figure 11.

\subsection{Na, Al and alpha elements}

Na is synthesized during hydrostatic carbon-burning in massive stars and also through the NeNa cycle during
H-burning through the CNO-cycle in intermediate-mass and massive stars (Woosley \& Weaver 1995; Denisenkov \& Denisenkova 1990). 
The chemical analysis of Sodium must be performed taking into account NLTE effects, these effects being greater for higher
equivalent widths and lower gravities (Gratton et al. 1999; Lind et al. 2011; Smiljanic 2012). In order to 
account for the NLTE effects in the Na abundances we used the corrections of Gratton et al. (1999). These corrections were typically 
smaller than 0.10 dex, with higher values ​​for giants with lower $\log g$ (stars 784 and 1534). With this 
NLTE correction the range in the abundance ratio [Na/Fe] for the red clump stars of Tombaugh 1 goes from 0.38 to 0.05 dex, with a 
mean value of 0.17 $\pm$ 0.06. Star 1534, classified as a field giant, showed the strongest NLTE effects  
with a correction of 0.22 dex, mainly due to its low surface gravity ($\log g$ = 2.0).
 
Chemical mixtures in the stellar interior can significantly modify the surface [Na/Fe] (e. g. Charbonnel \& Lagarde 2010). 
Comparing Tombaugh 1 with the models of Charbonnel \& Lagarde (2010), the mean cluster overabundance of [Na/Fe] = 0.17 among the giants 
is in excellent agreement with the values expected for models with thermohaline and rotation-induced mixing: 
[Na/Fe] = 0.18 for M = 2.0 M$_\odot$ and rotational velocities of 110 km\,s$^{-1}$ on the ZAMS. 
The range of [Na/Fe] among the giants of Tombaugh 1 could be explained by a range of rotation velocities among the 
stars in ZAMS which produced the giants (Charbonnel \& Lagarde 2010).

The production of Al, Mg, Si, Ca and Ti occurs mainly in massive stars whereas the production of the iron-peak elements 
is dominated by SN Type Ia (e. g. Woosley \& Weaver 1995; Iwamoto et al. 1999). Thus, the chemical ratio of Al and 
alpha-elements with Fe can give us important information about the SNIa and SNII contributions to the 
galactic components (bulge, disk and halo). The mean abundances of Mg, Si and Ca relative to Fe for Tombaugh 1 
show essentially solar values of $+$0.03$\pm$0.05, $+$0.01$\pm$0.07 and $+$0.01$\pm$0.03, respectively. In the case of 
[Ti/Fe], we found for Tombaugh 1 a slightly overabundant value relative to the 
sun with a mean of $+$0.11$\pm$0.04 dex. Our values ​​of [X/Fe] for alpha elements in Tombaugh 1 are consistent 
with the disk giants of Luck \& Heiter (2007) and Mishenina et al. (2006) and also with open clusters with 
similar metallicity of literature (Figure 12). The decay of the [X/Fe] ratio to alpha elements with
increasing of metallicity in the disk stars, as observed in the Figure 12, can be explained by the 
SNIa yields (Iwamoto et al. 1999), i.e. by high creation of Fe and low generation of alpha-elements. 
[Al/Fe] for Tombaugh 1 is similar to [Ti/Fe], with a mean of $+$0.15, in agreement with the chemical pattern 
of Al in the galactic disk (Figure 12). 

\subsection{Neutron-capture elements}

The elements Y, Ba, Ce and Nd are formed mainly in the stellar interior by slow neutron-capture 
process (s-process) during the asymptotic giant branch (AGB) phase and are transported to the stellar 
surface by the third dredge-up (Busso et al. 1999). In Tombaugh 1, the light s-process element,Y, has a near solar [X/Fe] mean
of $+$0.06 $\pm$ 0.04 dex while [X/Fe] for heavy s-process elements (Ba, Ce and Nd) shows an excess compared to the sun, with a
mean of $+$0.35 $\pm$ 0.03 for Ba, $+$0.25 $\pm$ 0.06 for Ce, and $+$0.37 $\pm$ 0.05 for Nd. The difference between 
light and heavy s-process elements is an indicator of s-process efficiency (e. g. Luck \& Bond 1981, 1991; 
Busso et al. 2001; Pereira et al. 2011), implying a high s-process efficiency for Tombaugh 1. Other open clusters 
exhibiting this same behavior include the Hyades (De Silva et al. 2006), Berkeley 18, Berkeley 21, Berkeley 22 
and Berkeley 32 (Yong et al. 2012), Ruprecht 4, Ruprecht 7, NGC 6192 and NGC 6404 (Mishenina et al. 2013), among others. 
The s-process efficiency is an important observational constraint to stellar evolutionary models
(e.g. Busso et al. 2001) and is affected by metallicity, stellar mass and rotational 
velocity (e.g. Lugaro et al. 2003; Herwig et al. 2003).

Abundance measurements for s-process elements from the literature are highly inhomogeneous and 
difficult to compare with our results due to the use of different absorption lines, atomic parameters, and analysis 
methods (see. e.g., Yong et al. 2012 for a detailed discussion). Nevertheless, our s-process abundances for 
Tombaugh 1 agree with published s-process abundances for open clusters, as shown in Figure 13. Only our neodymium
abundances show a slight overabundance with respect to open clusters and disk field giants from the literature.

\subsection{The peculiar Tombaugh 1 field Cepheid XZ CMa}

XZ CMa (star 806 in Table 4) is a short-period Cepheid (P=2$^{d}$.56, Caldwell \& Coulson, 1987) situated within of coronal
region of Tombaugh 1 but classified as not cluster member (see section 3.2). Three papers in the literature analyzed in
detail the Cepheid XZ CMa (Turner, 1983; Diethelm, 1990; Yong et al. 2006). Turner (1983) and Diethelm (1990) conducted a
photometric analysis of XZ CMa, while Yong et al. (2006) analyzed XZ CMa with a high-resolution spectroscopy. Turner (1983),
via UBV photoelectric photometry, defined XZ CMa as unlikely member of Tombaugh 1 and found that XZ CMa probably has an
unresolved blue companion which is aprox. 2.5 magnitudes fainter in V, due the phase of minimum in the U-V curve is shifted
from the phase of minimum light by roughly 0.2 to 0.3 of the star's period. Subsequently, based in Walraven VBLUW
photometric system, Diethelm (1990) derived the mean atmospheric parameters ($T_{\rm eff}$, $\log g$ and [Fe/H]) of XZ CMa,
obtaining $T_{\rm eff}$=6000 K, $log g$=2.3 (dex) and [Fe/H]=$-$0.50$\pm$0.10, with differences between our atmospheric
parameters and Diethelm (1990) values of $\Delta T_{\rm eff}$=0 K, $\Delta log g$=0.4 (dex) and $\Delta [Fe/H]$=0.03 (dex). 

Lastly, Yong et al. (2006) determined the atmospheric parameters and the chemical abundances of three alpha-elements (Si,
Ca and TiII) to Cepheid XZ CMa, using same method but different line-list that used in this work. Our atmospheric
parameters $T_{\rm eff}$, $\log g$ and $\xi$ exhibit different values from those found by Yong et al. (2006), with
differences of 750 K, 1.12 (dex) and 1.58 km\,s$^{-1}$, respectively. However, we and Yong et al. (2006) obtained similiar
values of metallicity to XZ CMa ($\Delta [Fe/H]$=0.04). Probably, the difference of $T_{\rm eff}$, $\log g$ and $\xi$ displayed
in this work and in Yong et al. (2006) is due to observation of Cepheid XZ CMa in distinct pulsation phase, which causes
the determination of different values of atmospheric parameters ($T_{\rm eff}$, $\log g$ and $\xi$) and similar
metallicity. Finally, in both studies an overabundance of alpha elements in XZ CMa was found, with mean of alpha elements
in our analysis of [$\alpha$/Fe]=0.13 and in Yong et al. (2006) of [$\alpha$/Fe]=0.21, characteristic of Cepheid stars in the
outer disk (e.g., see Fig 15 of Yong et al. 2006).

Our results show that Cepheid XZ CMa has a chemical pattern similar to that presented by disk field stars and open clusters
(see Fig. 11 and 12). However, in Figure 13 we note that the Tombaugh 1 field star XZ CMa exhibits a high overabundance of
Ba compared with field giants from literature. To demonstrate the high Ba abundance in this star, in Figure 14 we present
the observed and synthetic spectra in the region around the absorption line of Ba II 5853 \AA{}. Classical Cepheids, like
XZ CMa, are not expected to present a high overabundance of s-process elements, as Ba, since such stars not evolved to AGB;
e.g., cepheids FO Cas, EW Aur, EE Mon and FF Aur with similar metallicity of XZ CMa presents the ratio [Ba/Fe] of 0.17,
0.24, 0.03 and 0.13, respectively (Andrievsky et al., 2014). The chemical abundances of Ba in disk Cepheids is known to
suffer from NLTE effects (Andrievsky et al. 2013; Andrievsky et al. 2014). However, the NLTE correction for Ba II line 5853
\AA{} is not especially large, averaging around $-$0.1 dex (Andrievsky et al. 2013), does not having any significant effect
in the high overabundance obtained for XZ CMa. We will discuss the case of this star in the final section.

\section{Discussion and Conclusions}

In this paper we have presented the first study of Tombaugh 1 using both high-resolution spectroscopy and precision 
$uvbyCa$H$\beta$ photometry. Our results for the abundance ratios of elements from Na to Ni and the
cluster fundamental parameters of distance and age tag this open cluster as an intermediate-age (0.95 Gyr) cluster 
belonging to the galactic thin disk. As such, it allows the addition of one more data point to the census of star clusters
used to map the chemical history of the disk, falling within a galactocentric zone where there is universal agreement
that a significant change in mean metallicity occurs among all classes of objects populating the thin disk. Where
disagreement arises is in the exact form and location of the transition region. Does Tombaugh 1 lie along a uniform linear 
gradient extending from R$_{GC}$ = 5 kpc to 20 kpc, or does the gradient change slope beyond the solar circle? If it changes, where
does the transition occur and why? The growing evidence from studies of distant anticenter open clusters and Cepheids
(e.g. Magrini et al. (2009); Lepine et al. (2011); Yong et al. (2012); Korotin et al. (2014), among others) is that 
the metallicity gradient beyond R$_{GC}$ = 13 kpc is considerably flatter than that between 9 and 13 kpc (see Fig. 15). 

In Figure 15 we show the radial metallicity gradient from Magrini et al. (2009) (blue points), with the addition of our 
spectroscopic results for Tombaugh 1 (red point) with [Fe/H] =$-$0.02$ \pm$ 0.05 and R$_{GC}$ = 10.46 kpc. If we use 
the lower photometric value of [Fe/H] =$-$0.10,
R$_{GC}$ would be reduced to 10.36 kpc, a negligible shift in distance on this scale. Also plotted are additional open clusters 
analyzed with high-resolution spectroscopy (green squares): IC 4725 and NGC 6087 (Gratton 2000); NGC 6603, NGC 2539, NGC 2447, IC 2714 
and NGC 5822 (Santos et al. 2009); NGC 6192, NGC 6404 and NGC 6583 (Magrini et al. 2010); 
NGC 7160 (Monroe \& Pilachowski 2010); Cr 110, NGC 2099, NGC 2420 and NGC 7789 (Pancino et al. 2010); 
Tombaugh 2 (Villanova et al. 2010); NGC 3114 (Santrich et al. 2013); NGC 4815 and NGC 6705 (Magrini et al. 2014); 
NGC 4337 (Carraro et al. 2014d); Trumpler 20 (Carraro et al. 2014b). The use of the
spectroscopic value alone is tied to an apparent offset between the photometric abundance scale, for Str\"omgren photometry
tied to high dispersion spectroscopy of F dwarfs, and the red giant high-dispersion spectroscopic scale, often distantly coupled to the sun. The issue is
apparent in Fig. 15 where, inside R$_{GC}$ = 9.8 kpc, no cluster has [Fe/H] below -0.1 and, more important, even ignoring the
super-metal-rich outliers, the typical cluster [Fe/H] at all ages is $+$0.1. While a virtually identical pattern was found
by Twarog et al. (1997), the lower limit and mean abundances from photometry and medium-resolution spectroscopy
of cluster red giants
were [Fe/H] =$-$0.2 and 0.0, respectively. Similar offsets between spectroscopic abundances of red giants and the photometry of 
F dwarfs have been found in NGC 3680 (Anthony-Twarog et al. 2009), NGC 5822 (Carraro et al. 2011), NGC 6819 (Anthony-Twarog et al. 
2014), and NGC 752 (Twarog et al. 2015). In the cases of NGC 3680, NGC 6819, and NGC 752, high dispersion spectroscopic analysis
of the F dwarfs agrees with the photometric abundances. If this offset to the spectroscopic scale applies to 
giants across all metallicities, the trend in Fig. 15 remains correct, even if the curve is shifted vertically by 0.1 dex.

We observe that Tombaugh 1 is consistent with the trend defined by Magrini et al. (2009) for the metallicity gradient, 
with Tombaugh 1 located in the inner disk (R$_{GC}\lesssim$ 12 kpc). The existence of an apparent transition zone ranging
from R$_{GC}$ = 10 to 12 kpc between an inner and outer disk lends support to the contention that metallicity evolution in these 
two regions occurs in different ways (Magrini et al. 2009; Lepine et al. 2011). According to Lepine et al. (2011), this 
behavior is due to a barrier created by a void in the interstellar gas in the region of the corotation radius 
of the main spiral structure. This dynamical interaction produces an inward flow of the gas on the inside of the corotation
zone of the Galaxy but an outward flow in the outer disk regions.

In recent years the abundances of the s-process elements in open clusters have become a target of 
intense study (e.g. D'Orazi et al. 2009, 2012; Jacobson et al. 2011; Maiorca et al. 2011; 
Jacobson \& Friel 2013; Mishenina et al. 2013, 2015). This recent interest was sparked by the unexpected results
of D'Orazi et al. (2009) for a sample of twenty open clusters. D'Orazi et al. (2009) found that [Ba/Fe] increases 
as cluster age decreases, contrary to the predictions of yields for Ba from AGB stars 
(e. g. Travaglio et al. 1999; Busso et al. 2001). Later work supplied confirmation for other s-process elements from
unevolved stars in open clusters: Ba (Mishenina et al. 2013; Jacobson \& Friel 2013), Ba and La (Jacobson et al. 2011), 
and Y, Zr, La, and Ce (Maiorca et al. 2011). However, Jacobson \& Friel (2013) didn't find a trend for [X/Fe] for La and Zr
versus age for their sample of 19 open clusters, which could indicate that the source of the s-process 
abundance trend with age doesn't affect all s-process elements equally. Among field stars, some s-process elements, Zr 
(Reddy et al. 2003) and Ba (Bensby et al. 2005), also show an increase in [X/Fe] with the decreasing age, 
while others, Y (Bensby et al. 2005), Ba and Ce (Reddy et al. 2003) do not. 

In this context our photometric and spectroscopic analysis classifies Tombaugh 1 as intermediate age (0.95 Gyr), 
with an enrichment of heavy s-process elements (Ba with $+$0.35 $\pm$ 0.03 dex, Ce with $+$0.25 $\pm$ 0.06 dex 
and Nd with $+$0.37 $\pm$ 0.05 dex) and solar values to Y ($+$0.06 $\pm$ 0.04), indicating a high
efficiency in the synthesis of the s-process elements. Some open clusters with similar ages show 
enrichment of the s-process elements similar to that found for Tombaugh 1, e.g. NGC 5822 (0.9 Gyr)
(Carraro et al. 2011) and NGC 3680 (1.7 Gyr) (Anthony-Twarog et al. 2009) with [Ce/Fe] = 0.25 and 
[Ce/Fe] = 0.26, respectively (Maiorca et al. 2011). 

The reason why open clusters younger than $\sim$1.5 Gyr (Maiorca et al. 2011) contain an overabundance of some s-process elements
(mainly Ba) compared to the old open clusters still isn't understood. D'Orazi et al. (2009) and Maiorca et al. (2011) have proposed 
a scenario with models of extra-mixing phenomena with high efficiency in the production of the neutron 
source $^{13}$C in stars with M $\leq$ 1.5M$_\odot$ (Busso et al. 2007; Nordhaus et al. 2008; Trippella et al. 2014; Nucci \& Busso 2014). 
Very recently, Mishenina et al. (2015) suggested that the Ba overabundance in open clusters
could be due to action from the intermediate neutron-capture process, or i-process (Cowan \& Rose 1977). 
However, as Mishenina et al. (2015) pointed out, it remains difficult to know which open cluster stars would be the host of 
the i-process; low-metallicity stars are a more probable example of these hosts (Bertolli et al. 2013; Dardelet et al. 2015). 
Indeed, confirmation of the enrichment of s-process elements in young clusters requires 
the analysis of a large and homogeneous sample of young and old open clusters with 
well-determined s-process abundances. 

The low number of open clusters with both reliable 
photometric and spectroscopic parameters, about 13.2\% of the known open clusters 
as defined by the 2014 update of the Dias et al. (2002) catalog, is just one of the factors that hinder a definitive characterization 
of the galactic metallicity gradient, as well as its variation over time and azimuthally within the disk for individual
elements. Studies of other poorly known open clusters like Tombaugh 1 using high-resolution spectroscopy and precision 
photometry to define reliably all of the key parameters that influence plots like Figs. 11, 12, 13 and 15 remain the key 
to forward progress in disentangling the complex system
known as the galactic disk. The next step in this direction is being 
conducted by large surveys like Gaia-ESO mapping the chemistry of all the components of the Galaxy. 

Finally, the overabundance of barium in Cepheid XZ CMa can be explained by an enhancement of s-process elements in the
interstellar medium (ISM) which produced XZ CMa or by mass transfer in a multiple-star system. Yong et al. (2006) found an
enhancement of La for a Cepheid sample in the outer disk and suggested that asymptotic giant branch stars have contributed
to the chemical evolution of the outer Galactic disk. XZ CMa is situated at the beginning of the outer disk ($R_{GC}$=13
Kpc, Yong et al. 2006), which makes XZ CMa one of the cepheid candidates rich in s-process elements formed by this ISM
suggested by Yong et al. (2006).

In a binary system, like Ba and CH stars, the enrichment of Ba is a consequence of mass transfer through stellar winds or
through Roche-lobe overflow from an AGB star (now the white dwarf) to a less evolved companion. Turner (1983) suggested the
presence of an unresolved blue companion B star to the Cepheid XZ CMa. However, the enrichment of Ba indicates a white
dwarf companion to XZ CMa. Thus, we suggest that XZ CMa can belong to a binary system with a white dwarf or a triple system
comprising a white dwarf and a B star. About one-third of Galactic Cepheids are known to have companions, and about 44\% of
those have more than one companion (Evans et al. 2005). Recently, in the study of the occurrence of classical cepheids in
binary systems, Neilson et al. (2015) pointed out that a fraction of binary systems may evolve to a system composed of a
Cepheid with a white dwarf companion. Harris \& Welch (1989) commented that due the occurrence of mass transfer in binary
Cepheids an evolutionary connection between Ba stars and binary Cepheids would be possible. In addition, Gonzalez \&
Wallerstein (1996) found significant similarities between binaries Cepheids, and Ba and CH stars, as orbital parameters and
mass range.

UV observations of XZ CMa can be used to confirm its binarity and reveal the nature of its companion (e.g. Evans 1992). In
the case of a hot companion to XZ CMa like B main-sequence star suggested by Turner (1983), the presence of a strong Balmer
line, H$\epsilon$ (3970.07 \AA), in the Cepheid spectrum also can be interpreted as the signature of this blue companion
(Kovtyukh et al. 2015). Because of the wavelength coverage of our XZ CMa spectrum (4200 \AA to 9000 \AA) was not possible to
perform this investigation. The discovery of binaries Cepheids is important because unresolved companions is one of the
factors that contribute to the scatter around the ridge-line period-luminosity relationship (Szabados \& Klagyivik, 2012).
In particular, the detection of a Cepheid-white-dwarf binary will give important constraint regarding the most massive 
progenitors of white dwarfs (Landsman et al. 1996). 



\acknowledgments

Extensive use was made of the WEBDA database maintained by E. Paunzen at the University of Vienna, Austria (http://www.univie.ac.at/webda). 
The filters used in the program were obtained by BJAT and BAT through NSF grant AST-0321247 to the University of Kansas. NSF support for 
this project was provided to BJAT and BAT through NSF grant AST-1211621.
J.V. Sales Silva acknowledges the support provided by CNPq/Brazil Science without Borders program (project No. 249122/2013-8). C. Moni Bidin
acknowledges support by the Fondo Nacional de Investigaci\'on Cient\'{\i}fica y Tecnol\'ogica (Fondecyt), project No. 1150060. E. Costa 
acknowledges support by the Fondo Nacional de Investigaci\'on Cient\'{\i}fica y Tecnol\'ogica (projecto No. 1110100, Fondecyt) and the Chilean Centro de Excelencia en Astrof\'{\i}sica y
Tecnolog\'{\i}as Afines (PFB 06).



{\it Facilities:} \facility{Magellan: Baade (IMACS)}, \facility{CTIO:1.0m (Y4KCam)}.

\clearpage



\begin{figure}
\centering
\includegraphics[width=\columnwidth]{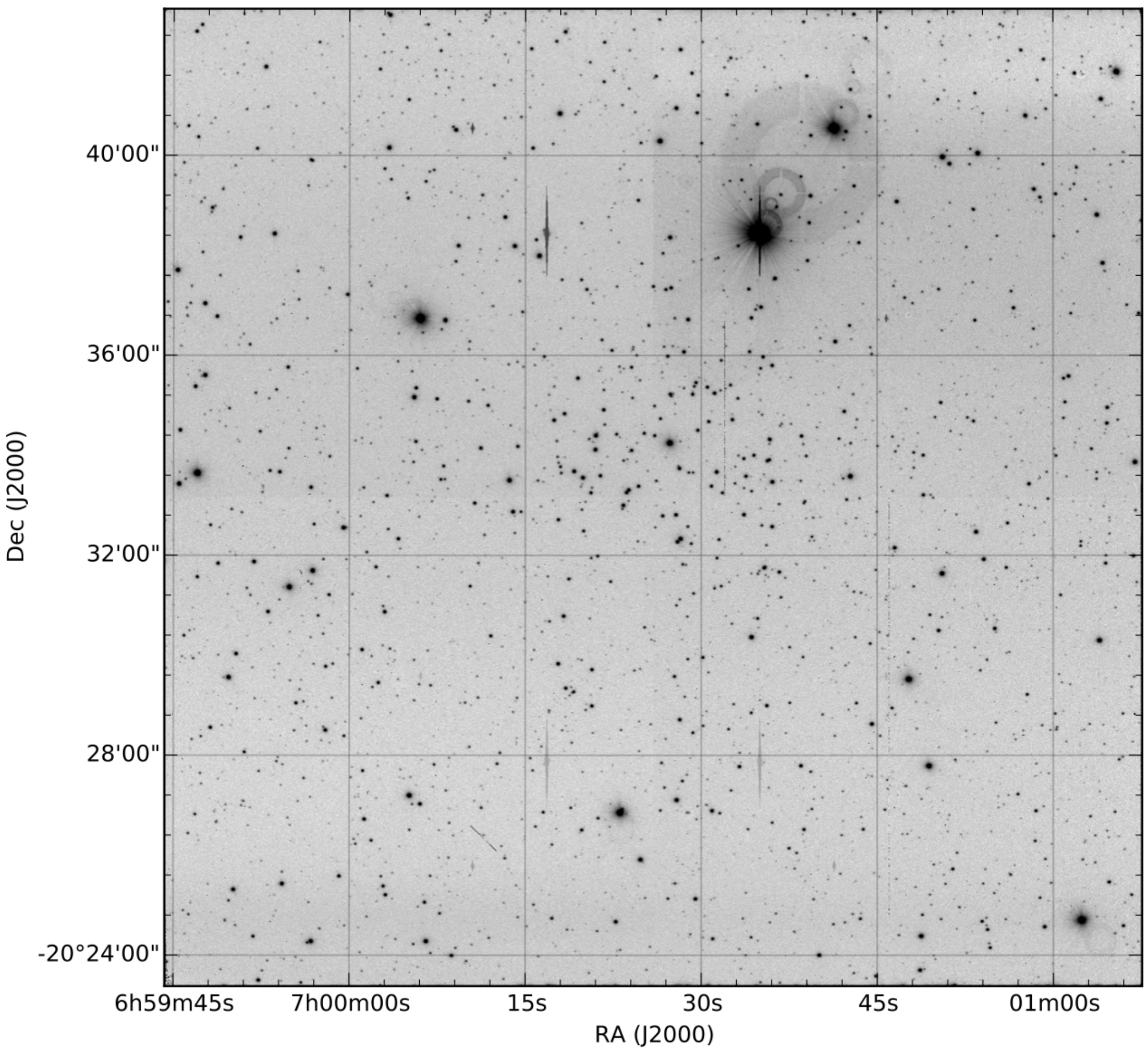}
\caption{A 1200 sec exposure in the $v$ filter. North is down and East to the right. The field is 20$^{\prime}$ on a side.}
\end{figure}

\begin{figure}
\centering
\includegraphics[width=\columnwidth]{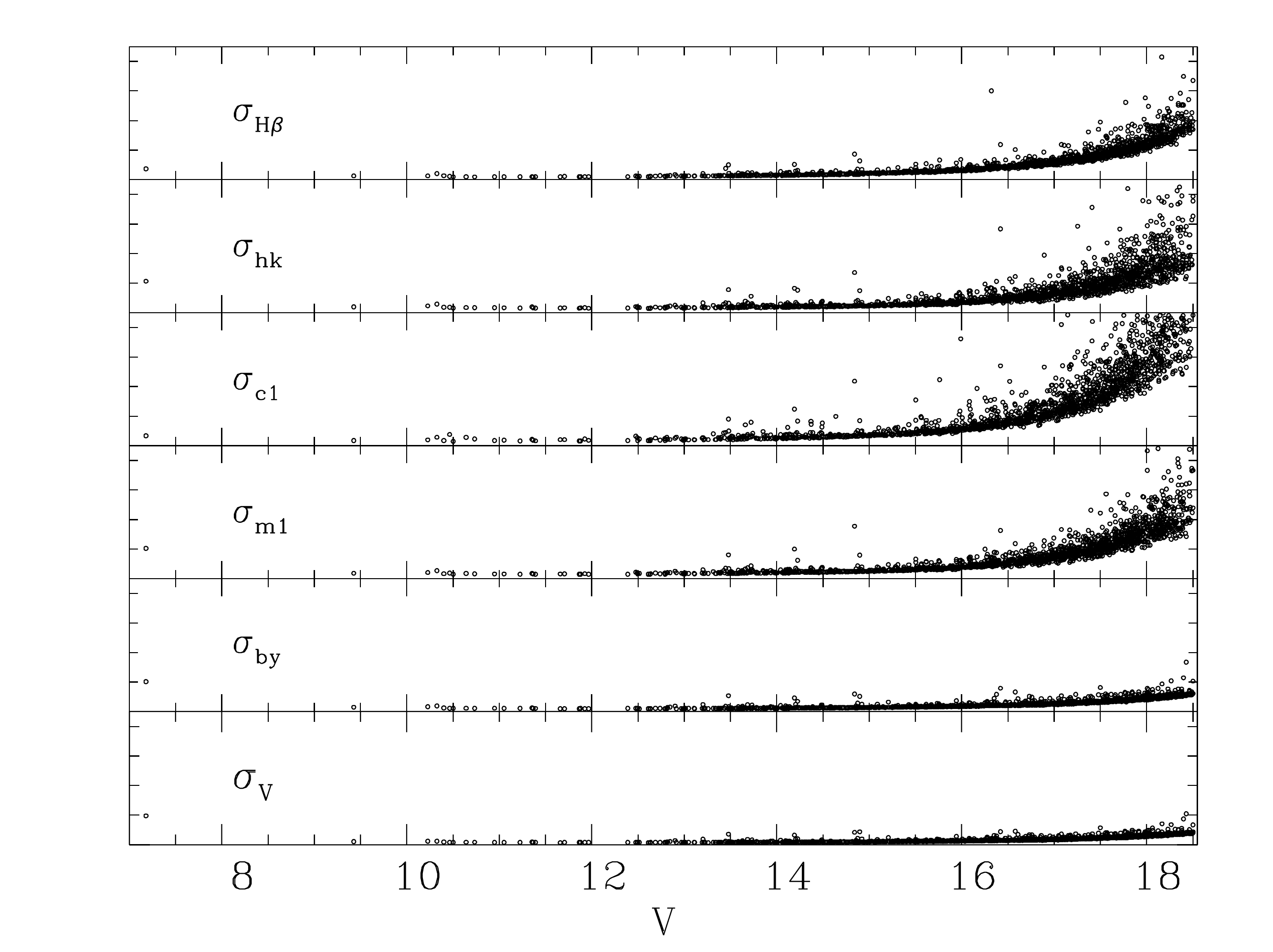}
\caption{Trend of global photometric errors in magnitude and colors as a function of $V$ magnitude. See text for details.}
\end{figure}

\begin{figure}
\centering
\includegraphics[width=\columnwidth]{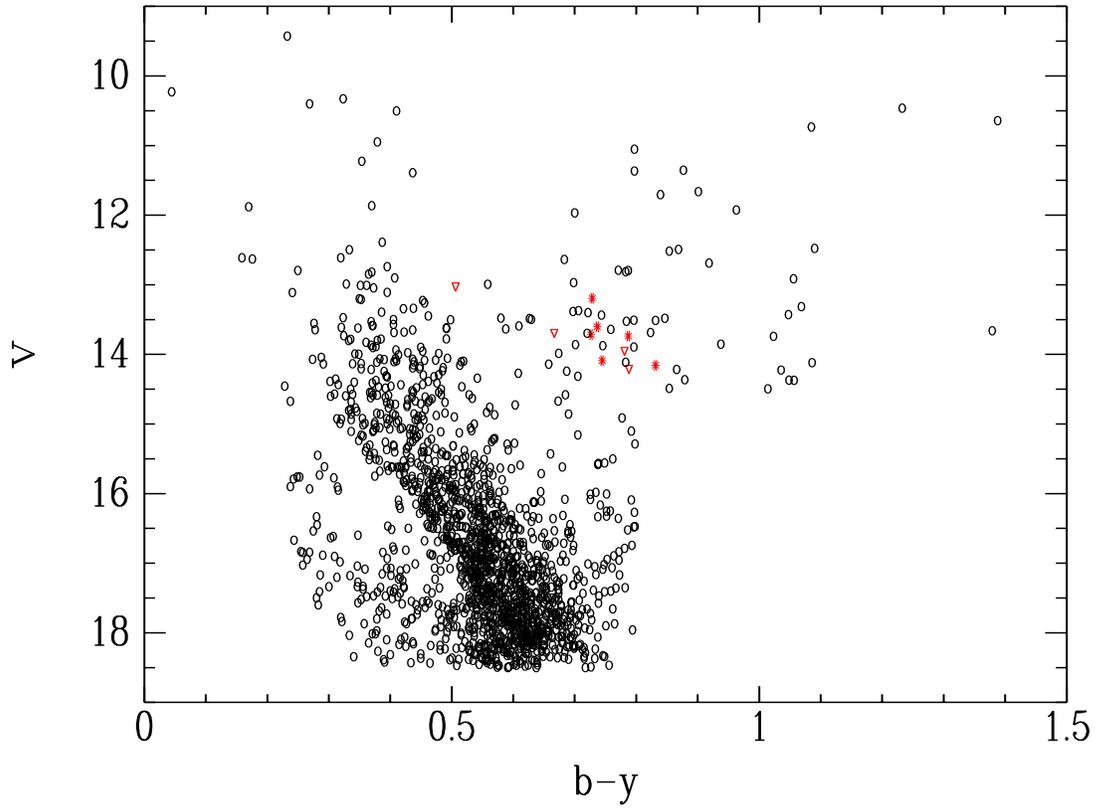}
\caption{Color-magnitude diagram of Tombaugh~1. Red symbols identify the ten potential cluster stars observed with IMACS. 
Starred points are probable members, while open triangles are non-members. See text for details.}%
\end{figure}

\begin{figure}
\centering
\includegraphics[width=\columnwidth]{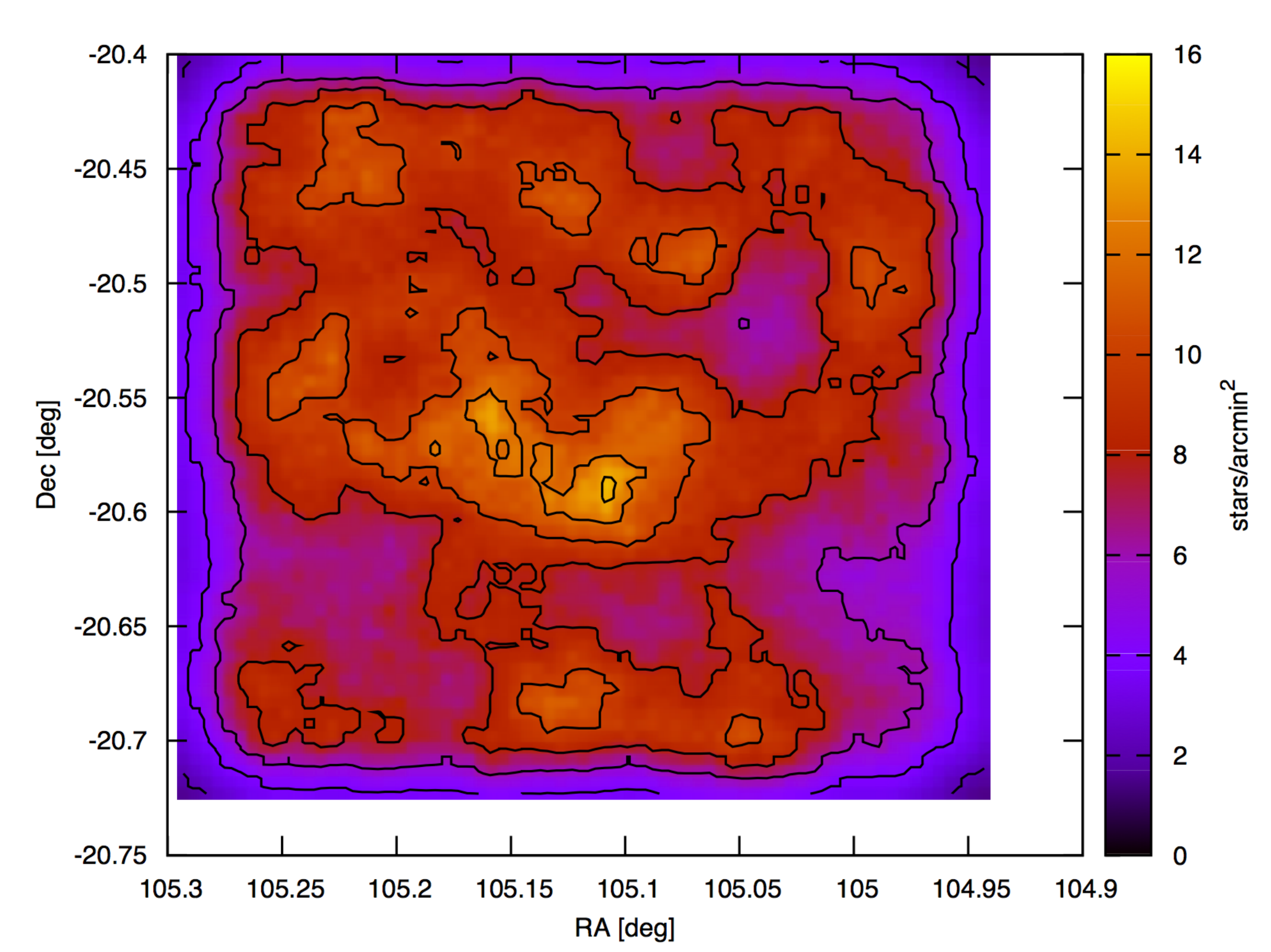}
\caption{Density contour map for Tombaugh 1 field. North is up, East to the left, and the field corresponds to 20$^{\prime}\times$20$^{\prime}$ on the sky}%
\end{figure}

\begin{figure}
\centering
\includegraphics[width=\columnwidth]{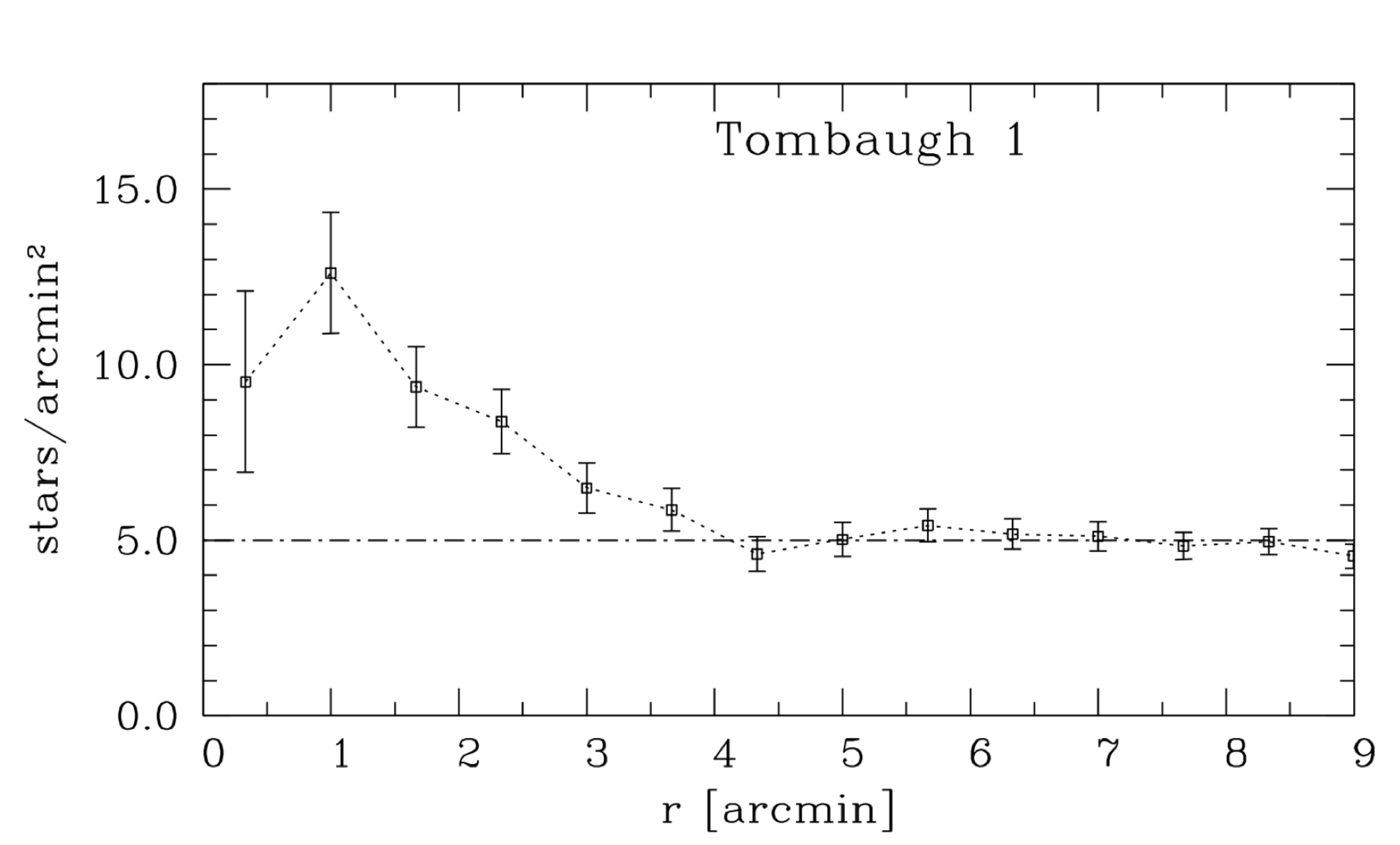}
\caption{Radial surface density profile. To define concentric rings, we used the nominal cluster center.}
\end{figure}

\begin{figure}
\centering
\includegraphics[width=\columnwidth]{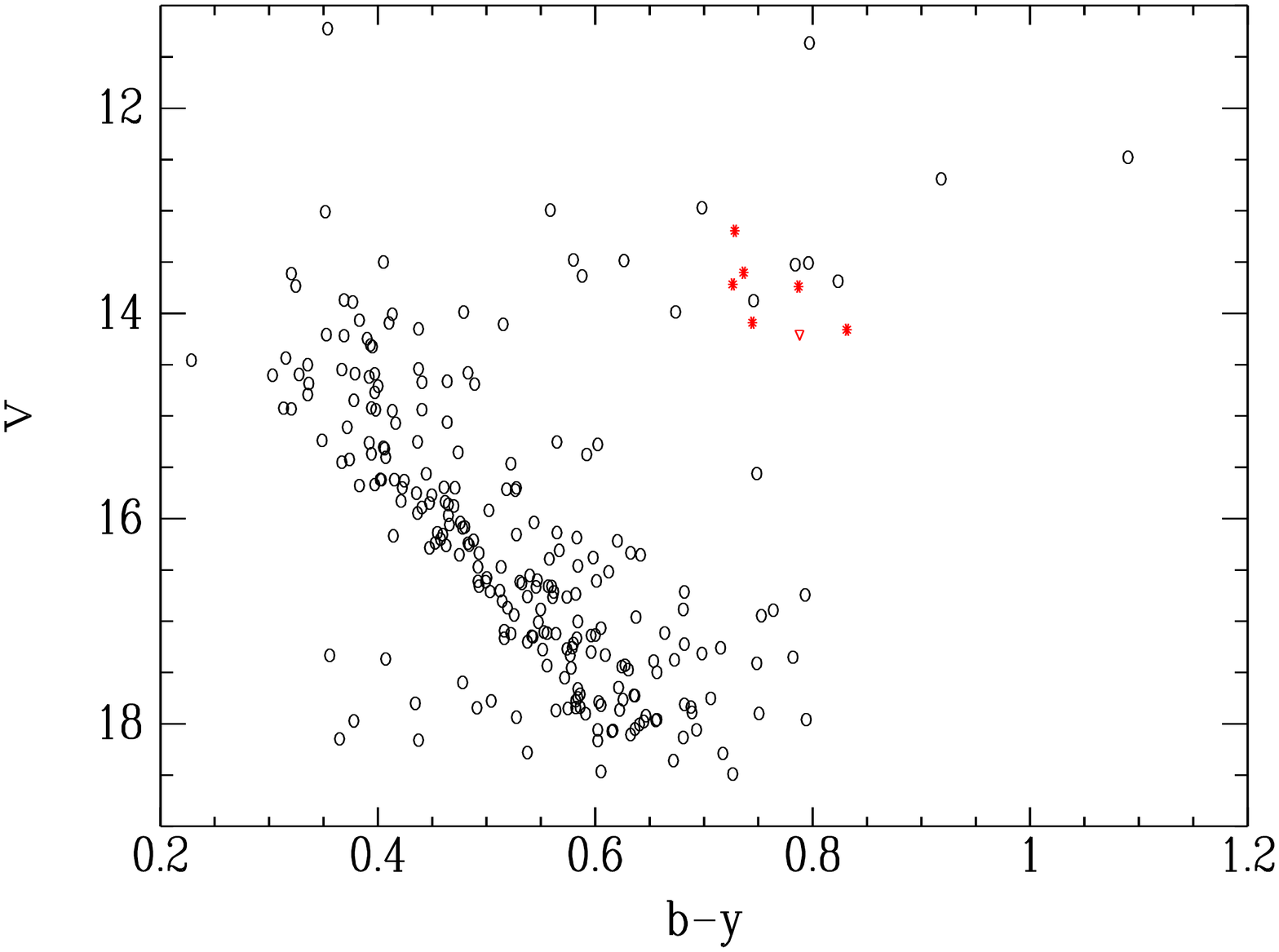}
\caption{CMD of the cluster within 3.5$^{\prime}$ of the cluster center. Symbols have the same meaning as in Fig. 3.}
\end{figure}

\begin{figure}
\centering
\includegraphics[width=\columnwidth]{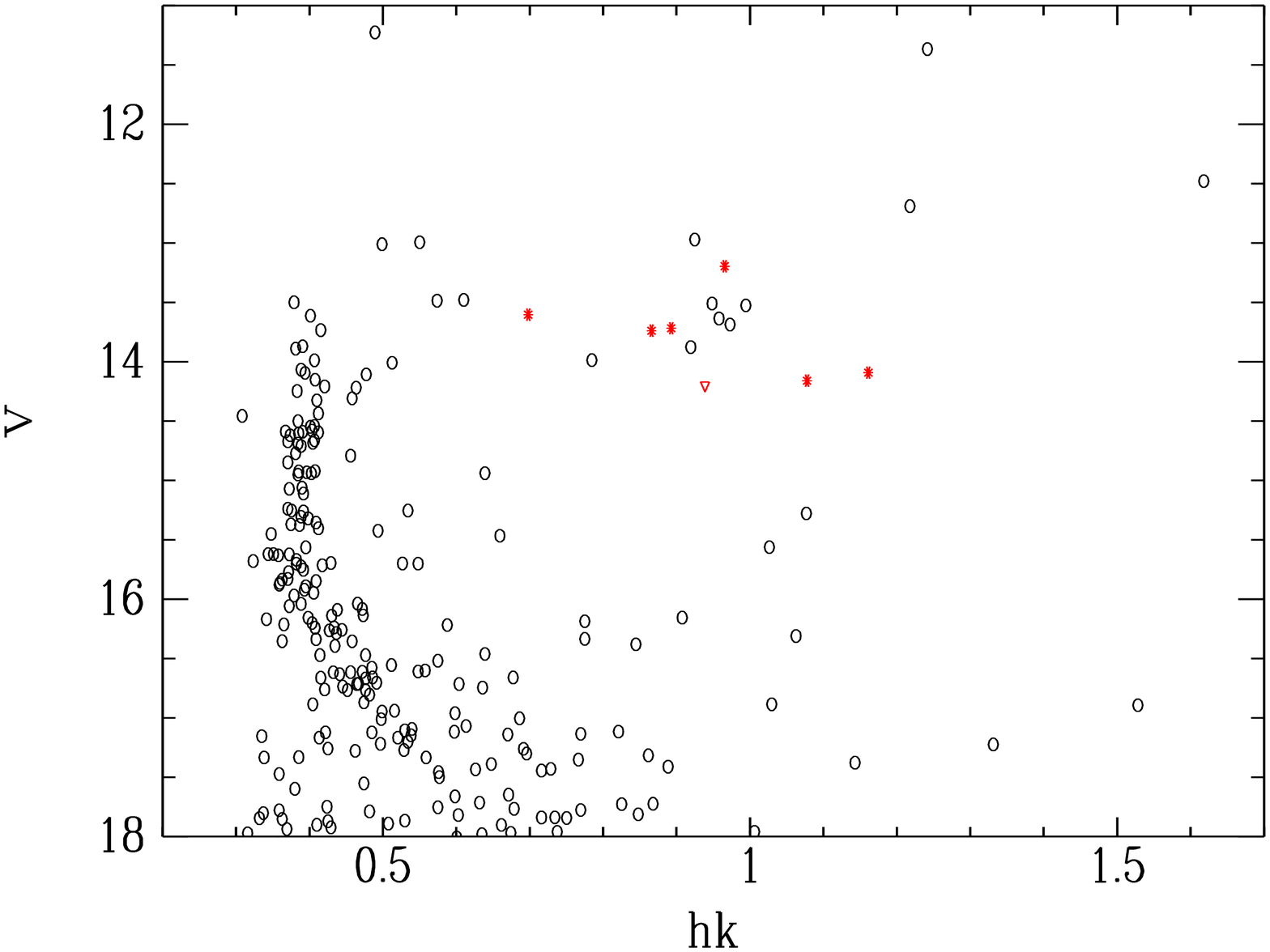}
\caption{$V, hk$ CMD of the cluster within 3.5$^{\prime}$ of the cluster center. Symbols have the same meaning as in Fig. 6.}
\end{figure}

\begin{figure}
\centering
\includegraphics[width=\columnwidth]{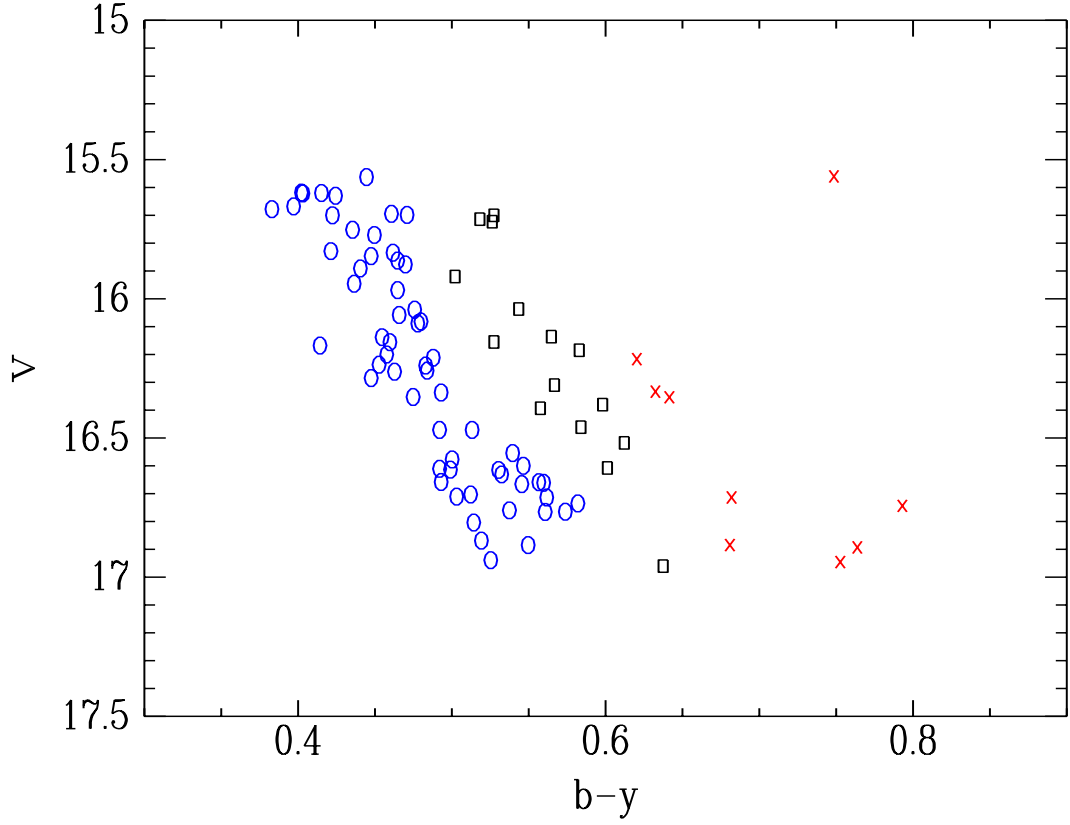}
\caption{CMD of the cluster unevolved main sequence within 3.5$^{\prime}$ of the cluster center. Blue open circles are probable single star members, black squares are potential binaries, and red crosses are likely nonmembers.}
\end{figure}

\begin{figure}
\centering
\includegraphics[width=\columnwidth]{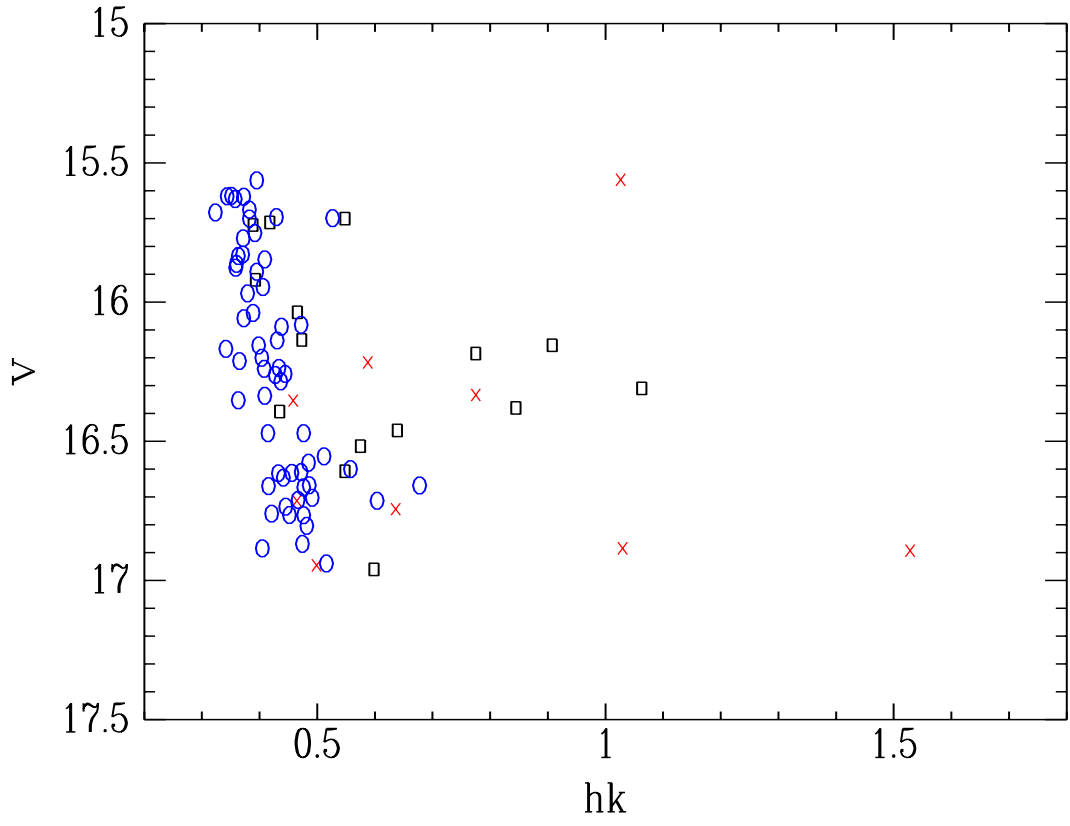}
\caption{Same as Fig. 8 using $hk$ as the temperature index.}
\end{figure}

\begin{figure}
\centering
\includegraphics[width=\columnwidth]{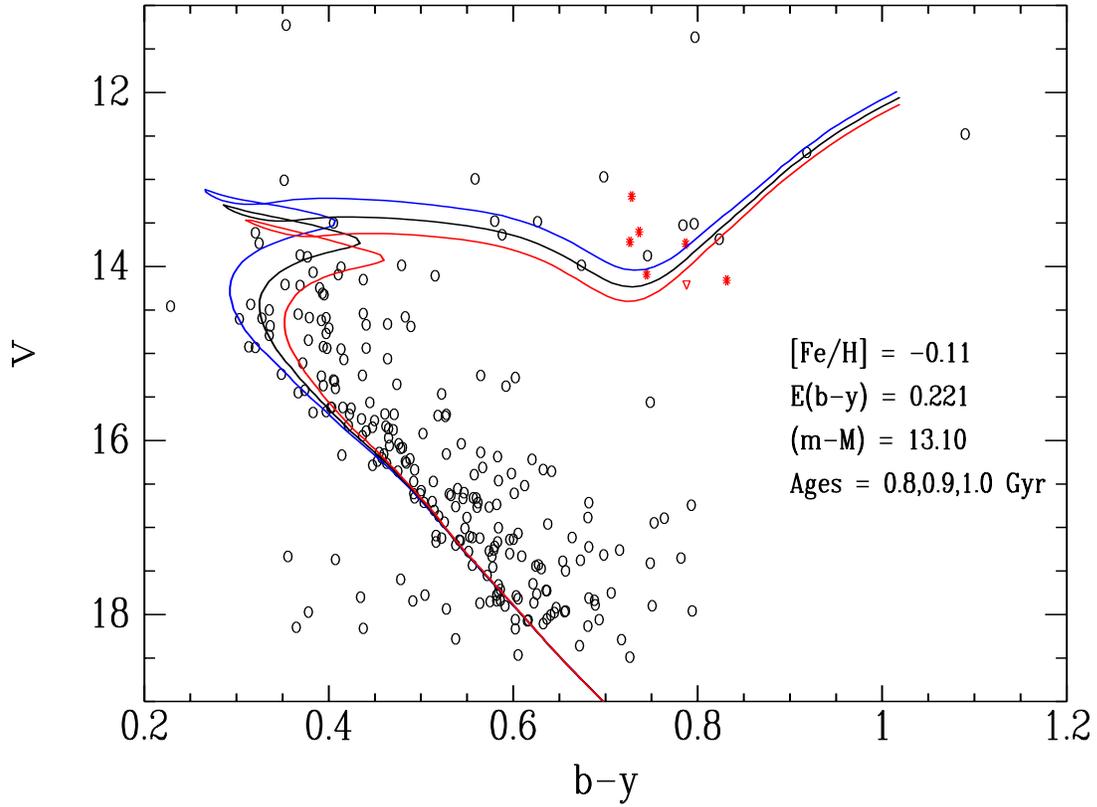}
\caption{CMD of Fig. 6 superposed on the VR isochrones with [Fe/H] = $-$0.11, assuming $E(B-V)$ = 0.303 and $(m-M)$ = 13.10. The isochrones have ages of 0.8 Gyr (blue), 0.9 Gyr (black) and 1.0 Gyr (red).}
\end{figure}

\begin{figure}
\label{alField}
\centering
\includegraphics[width=\columnwidth]{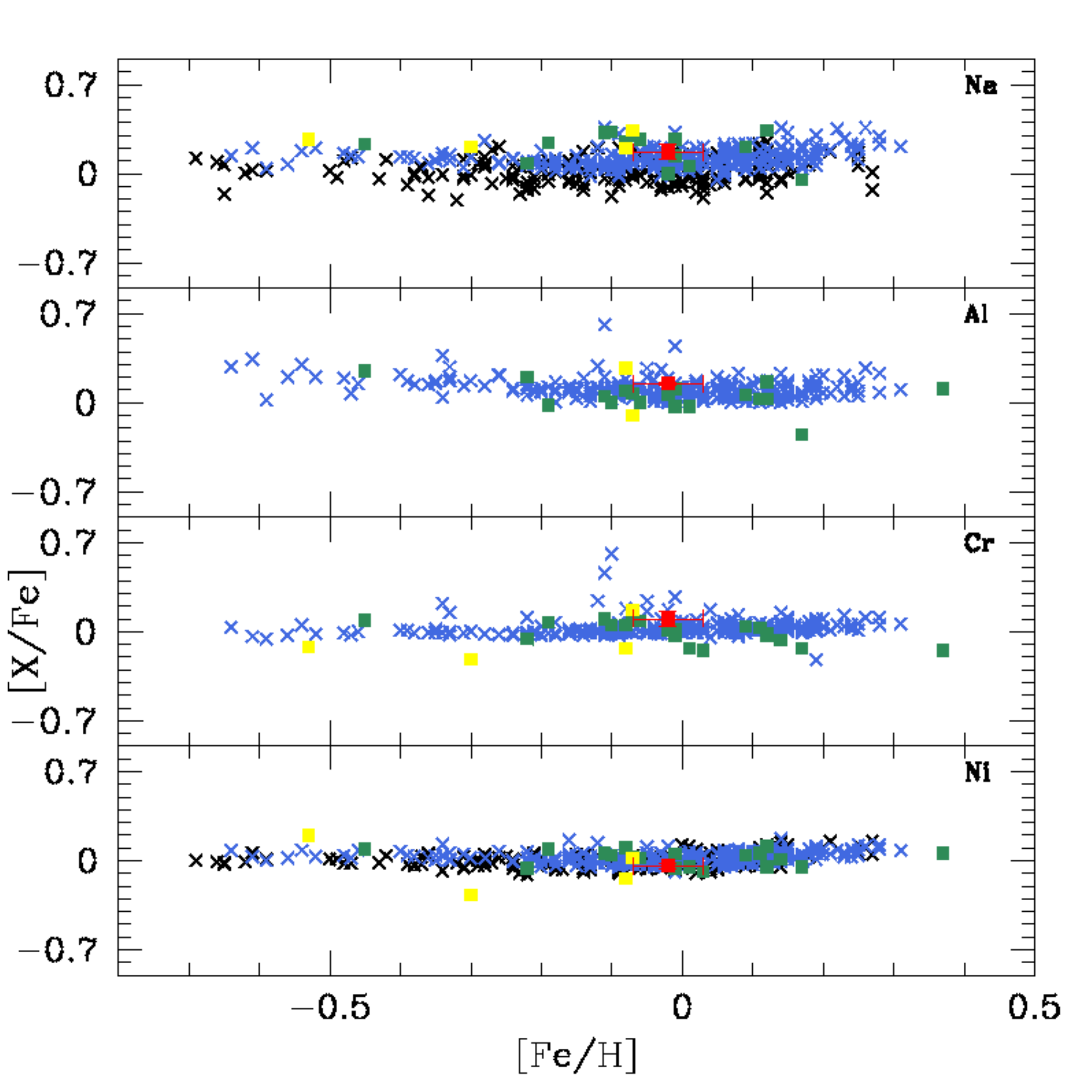}
\caption{Abundance ratios [X/Fe] vs. [Fe/H]. Blue crosses: field giants of Luck \& Heiter (2007); Black crosses: clump giants of Mishenina et al. (2006); Yellow squares: our sample of field giants stars; Red square: our mean abundances of Tombaugh 1; Green squares: open clusters from literature (NGC 6192, NGC 6404 and NGC 6583 of Magrini et al. 2010; NGC 3114 of Santrich et al. 2013; NGC 2527, NGC 2682, NGC 2482, NGC 2539, NGC 2335, NGC 2251 and NGC 2266 of Reddy et al. 2013; NGC 4337 of Carraro et al. 2014d; Trumpler 20 of Carraro et al. 2014b; NGC 4815 and NGC 6705 of Magrini et al. 2014; Cr 110, Cr 261, NGC 2477, NGC 2506 and NGC 5822 of Mishenina et al. 2015.}%
\end{figure}

\begin{figure}
\label{singlealphaField}
\centering
\includegraphics[width=\columnwidth]{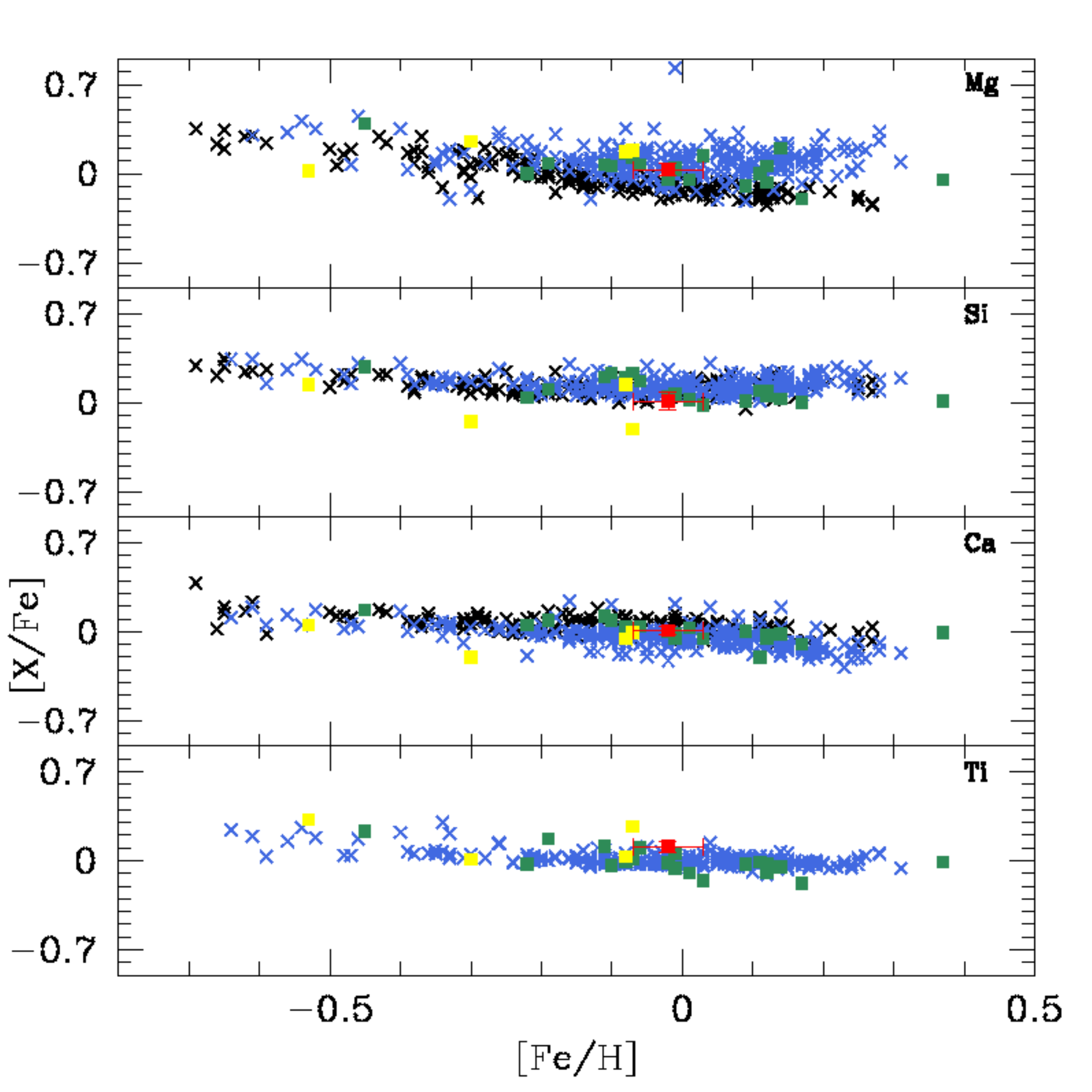}
\caption{Abundance ratios [X/Fe] vs. [Fe/H]. Symbols have the same meaning as in Figure 11.}%
\end{figure}

\begin{figure}
\label{sprocessTomb1}
\centering
\includegraphics[width=\columnwidth]{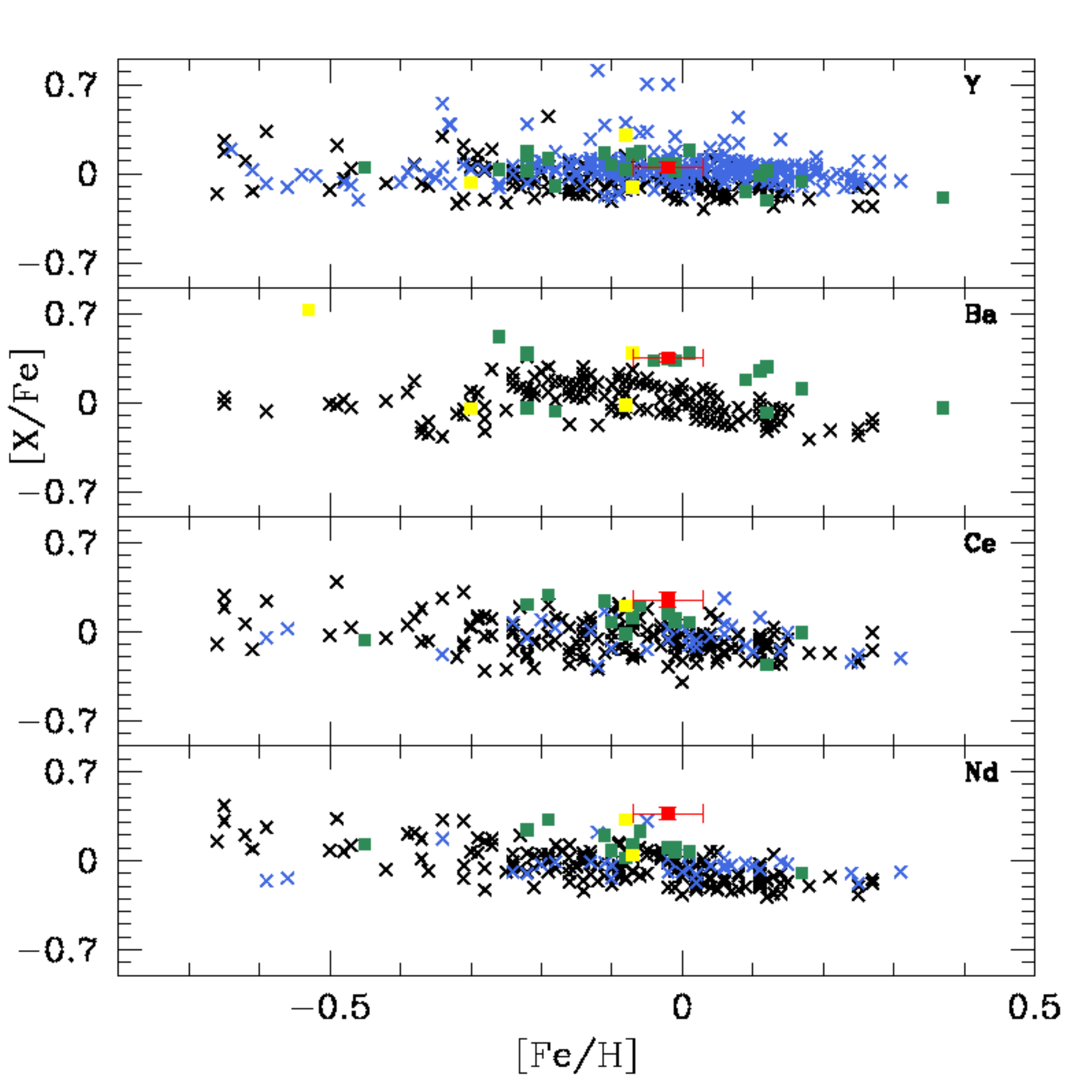}
\caption{Abundance ratios [X/Fe] vs. [Fe/H]. Symbols have the same meaning as in Figure 11. In the Y and Ba panel, we added the results of Mishenina et al. (2013) of the open clusters Berkeley 25, Berkeley 73, Berkeley 75, Ruprecht 4, Ruprecht 7, NGC 6192, NGC 6404 and NGC 6583 (green squares). One field star of our sample (806) exhibits Ba enrichment.}%
\end{figure}

\begin{figure}
\centering
\label{O6300-final}
\includegraphics[width=\columnwidth]{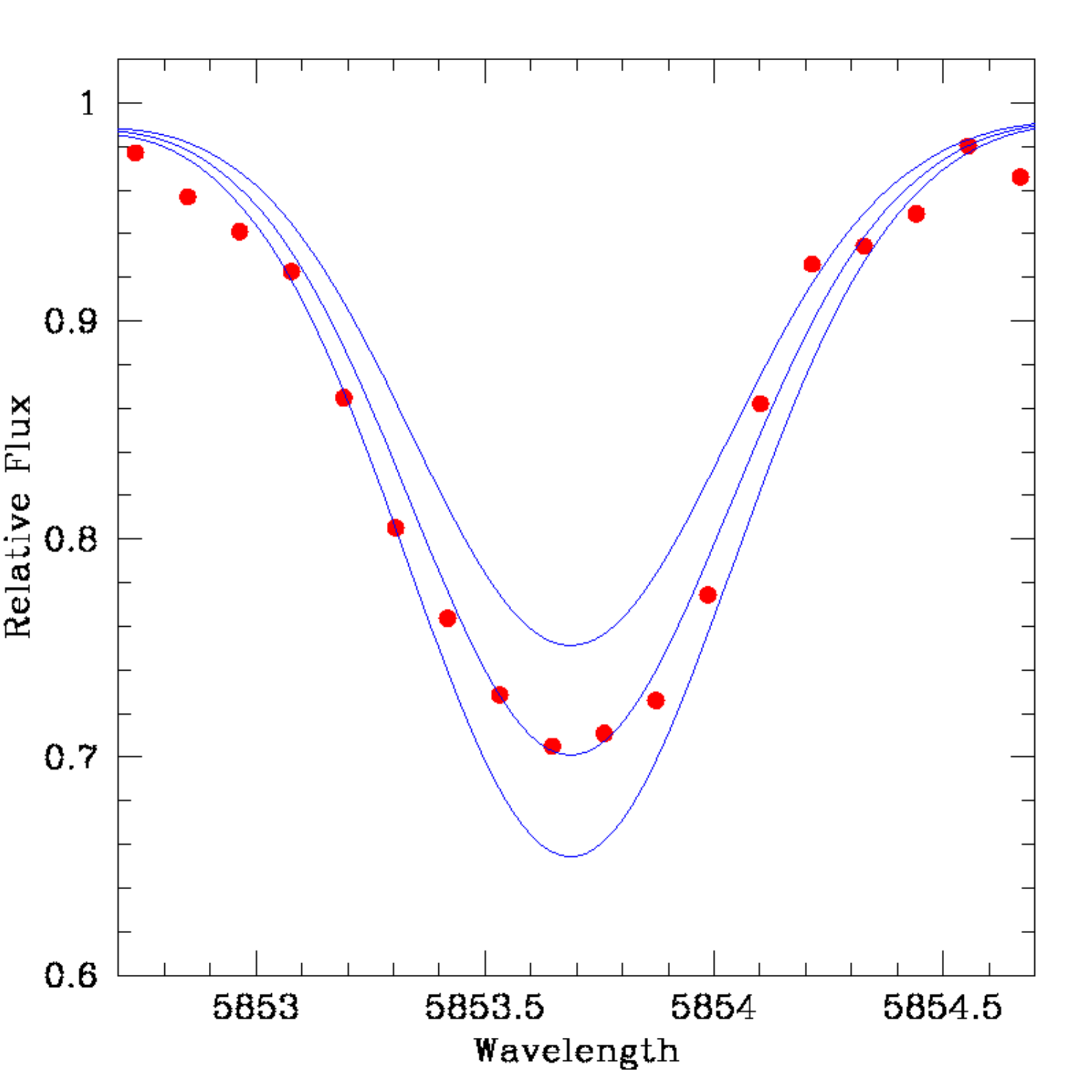}
\caption{Observed (dotted red line) and synthetic spectra (solid blue lines) in the region of the Ba II line at 5853 \AA{} for the field giant star 806. The synthetic spectra were calculated with the barium ratio abundances of [Ba/Fe]= 0.33, 0.73 and 1.13.}%
\end{figure}

\begin{figure}
\centering
\label{FeHxRgc2}
\includegraphics[width=\columnwidth]{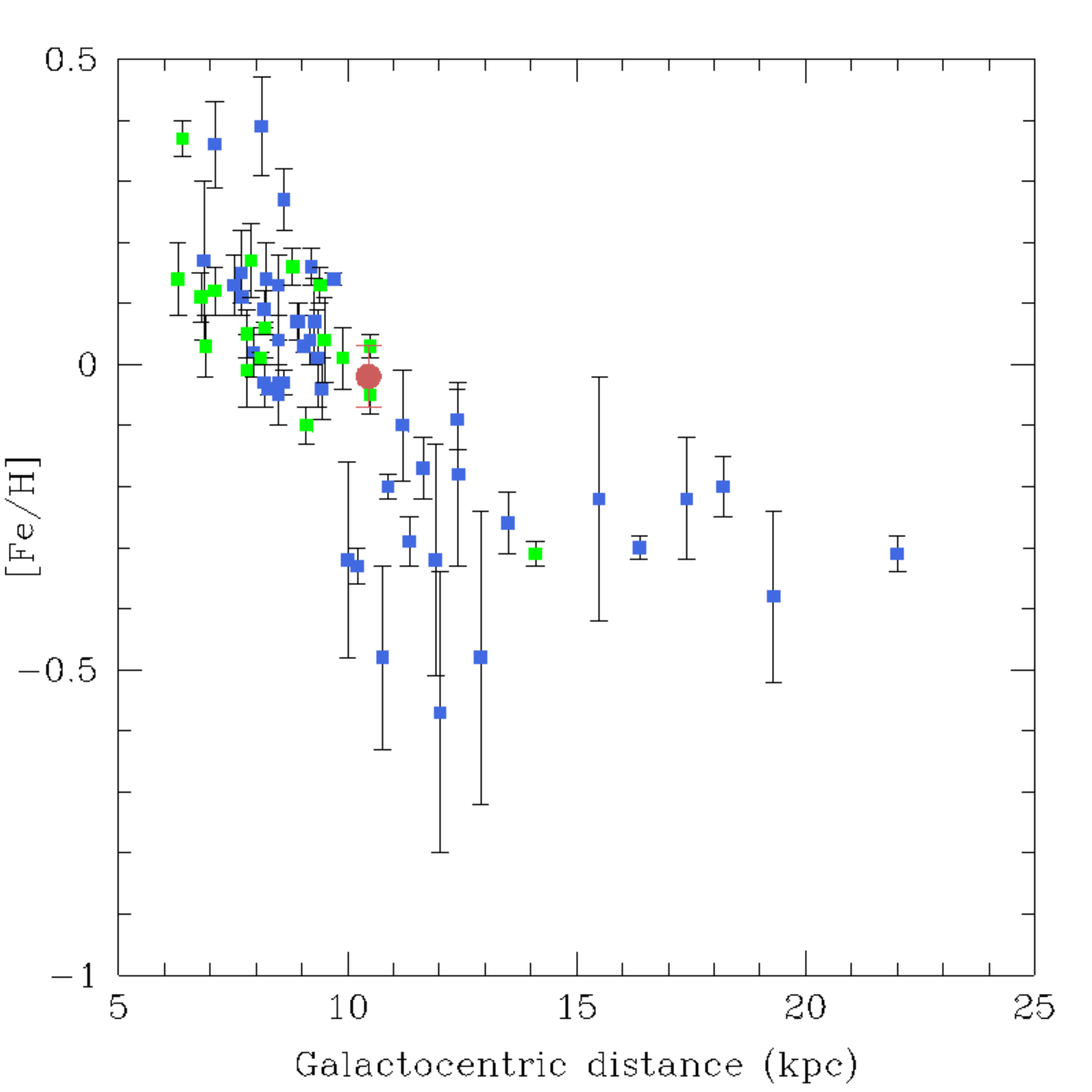}
\caption{Radial metallicity gradient from Magrini et al. (2009, blue squares) with the addition of Tombaugh 1 (red circle) and additional open clusters analyzed with high-resolution spectroscopy (green squares). See text for details. $R_{GC\odot}$ = 8.5 kpc.}%
\end{figure}

\clearpage







\clearpage

\begin{table}
\tabcolsep 0.1truecm
\caption{Str\"omgren photometric observations of Tombaugh~1}
\scriptsize
\begin{tabular}{lcccc}
\tableline
\noalign{\smallskip}
Date & Field & Filter & Exposures (sec) & airmass (X)\\
\noalign{\smallskip}
\tableline\tableline
\noalign{\smallskip}
Dec  05, 2010  & Tombaugh~1  & \textit{y} & 60, 600           &1.03\\
                         &                        & \textit{b} & 60, 600          & 1.02\\
                         &                        & \textit{H$\beta_{wide}$} & 60, 600 & 1.01$-$1.02\\
                         &                        & \textit{Ca} & 120, 1200 & 1.01\\
                         &                        & \textit{H$\beta_{narrow}$} & 120,1200 & 1.02\\
\tableline
Dec 06, 2010  & Tombaugh~1 & \textit{y}  & 2x60, 900                              & 1.03 \\
                                  &             & \textit{b}  & 2x60,900       & 1.02\\
                                  &             & \textit{v} & 60, 900          & 1.01$-$1.03\\
                                  &             & \textit{Ca} & 120,1500 & 1.04-1.08\\
\tableline
Dec 07 , 2010  & Tombaugh~1  & \textit{u}   & 10, 300                             & 1.02\\
                                  &                     & \textit{b}  & 10, 60       & 1.01\\
                                  &                     & \textit{v}    & 10, 100                    & 1.01\\
                                  &                     & \textit{Ca}  &  10, 200                             &  1.01\\
                         &                        & \textit{H$\beta_{wide}$} & 10, 60 & 1.01$-$1.01\\
                         &                        & \textit{H$\beta_{narrow}$} & 10, 200 & 1.02\\
\tableline
Dec 08, 2010  & Tombaugh~1 & \textit{u}   & 2x20, 200 2000  & 1.02$-$1.03 \\
                                  &                     & \textit{v}  & 20, 90, 900                       & 1.01$-$1.03\\
                         &                        & \textit{H$\beta_{narrow}$} & 20, 150, 1500 & 1.05\\
\tableline
Dec 09, 2010  &  Tombaugh~1&\textit{y} & 10, 60, 120, 600                              & 1.46$-$1.51\\ 
                                  &                      &\textit{b}  & 60,180,900   & 1.35$-$1.39 \\
                                  &                       & \textit{v}  & 100, 200, 1200   & 1.25$-$1.28\\
                                  &                       & \textit{Ca}  & 100, 300,1800    & 1.04$-$1.05\\
                         &                        & \textit{H$\beta_{wide}$} & 100, 200, 1200 & 1.16$-$1.18\\
                         &                        & \textit{H$\beta_{narrow}$} & 10, 300, 1800 & 1.09$-$1.11\\
\tableline
Dec 09, 2010  &  Tombaugh~1&\textit{y} & 60, 600                              & 1.47$-$1.49\\ 
                                  &                      &\textit{b}  & 180, 900   & 1.38$-$1.40 \\
                                  &                       & \textit{v}  & 200, 900   & 1.30\\
                                  &                       & \textit{Ca}  & 300, 1500    & 1.07$-$1.08\\
                         &                        & \textit{H$\beta_{wide}$} & 200, 900 & 1.30\\
                         &                        & \textit{H$\beta_{narrow}$} & 300, 1500 & 1.13$-$1.14\\
                                  &               & \textit{u} &  400, 1800 & 1.03$-$1.04\\
\tableline
\noalign{\smallskip}
\tableline
\end{tabular}
\end{table}

\clearpage

\begin{table*}
\tabcolsep 0.3truecm
\caption{Calibration Equations Summary}
\begin{tabular}{crrr}
\tableline
Index & slope & color term & Std. deviation \\
\tableline
$V$     & 1.00   & 0.05    & 0.010 \\
$b-y$   & 1.01   & --- & 0.002 \\
$hk$    & 1.07   & --- & 0.009 \\
H$\beta$& 1.18   & --- & 0.015 \\
$m_1$   & 0.92   & $-$0.075  & 0.025 \\
$c_1$   & 1.06  & --- & 0.021 \\
\tableline
\end{tabular}
\end{table*}  

\clearpage

\begin{landscape}
\begin{table*}
\tabcolsep 0.1truecm
\caption{Excerpt of the photometric catalog of Tombaugh~1.}
\begin{tabular}{llrccccccccccc}
\tableline
$\alpha(2000)$ & $\delta(2000)$ &  $V$ & $b-y$ & $hk$ & $H\beta$ & $m_1$ & $c_1$ & $\sigma_V$ & $\sigma_{by}$ & $\sigma_{m1}$ &  $\sigma_{c1}$ & $\sigma_{hk}$ & $\sigma_{\beta}$ \\
\tableline
   105.1726&  $-$20.6754&     7.179&   0.826&   1.377&   2.532&   0.453&   0.187&    0.0195&   0.0204&   0.0205&   0.0067&   0.0213&   0.0074\\
   105.0252&  $-$20.6121&     9.426&   0.232&   0.326&   2.769&   0.119&   0.793&    0.0021&   0.0030&   0.0037&   0.0035&   0.0040&   0.0026\\
   105.2605&  $-$20.4113&   10.227&   0.044&   0.234&   2.915&   0.145&   1.012&    0.0023&   0.0034&   0.0042&   0.0039&   0.0046&   0.0026\\
   105.1471&  $-$20.6414&   10.326&   0.323&   0.550&   2.673&   0.158&   0.406&    0.0023&   0.0039&   0.0055&   0.0057&   0.0058&   0.0042\\
   105.0961&  $-$20.4473&   10.400&   0.268&   0.473&   2.716&   0.209&   0.613&    0.0018&   0.0026&   0.0034&   0.0034&   0.0037&   0.0027\\
   105.2796&  $-$20.5640&   10.463&   1.232&   1.976&   2.574&   0.661&  -0.319&    0.0016&   0.0023&   0.0038&   0.0076&   0.0037&   0.0022\\
   104.9458&  $-$20.5607&   10.502&   0.410&   0.585&   2.582&   0.167&   0.251&    0.0016&   0.0023&   0.0030&   0.0030&   0.0032&   0.0019\\
   105.0569&  $-$20.5582&   10.641&   1.387&   1.216&   2.754&   0.050&   0.134&    0.0017&   0.0024&   0.0035&   0.0056&   0.0035&   0.0020\\
   105.0750&  $-$20.6804&   10.733&   1.084&   1.836&   2.550&   0.639&  -0.147&    0.0017&   0.0023&   0.0033&   0.0046&   0.0034&   0.0019\\
   105.1990&  $-$20.4917&   10.947&   0.379&   0.568&   2.620&   0.134&   0.411&    0.0016&   0.0024&   0.0032&   0.0035&   0.0034&   0.0020\\
   105.0211&  $-$20.4531&   11.051&   0.797&   1.338&   2.550&   0.471&   0.199&    0.0017&   0.0023&   0.0032&   0.0037&   0.0033&   0.0020\\
\tableline
\end{tabular}
\end{table*} 
\end{landscape}

\clearpage

\begin{landscape}
\begin{table*}
\tabcolsep 0.2truecm
\caption{Potential cluster stars observed with IMACS in the field of Tombaugh 1.}
\label{sample}
\begin{tabular}{lcccccccccc}\\\tableline\tableline
ID   &  RA(2000.0)  &   DEC(2000.0)   &    $V$     &   $b-y$  &    $V-I$ & RV$_{1}$      &       RV$_{2}$    & $\langle$RV$\rangle$ &    vsini      &  S/N \\\tableline
     &    hh:mm:ss  &    dd:mm:ss     &    mag     &    mag   &    mag   & (km s$^{-1}$) &     (km s$^{-1}$) & (km s$^{-1}$)        &  (km s$^{-1}$)&      \\\tableline

395  &$07:00:05.7$  & $-20:35:20.5$   &  $13.72$   &  $0.73$ &$1.19$& 81.6$\pm$1.8     &  82.7$\pm$2.4       &  82.1$\pm$0.7        &    $<$2.7     &  100 \\
663  &$07:00:18.7$  & $-20:31:31.0$   &  $14.09$   &  $0.75$ &$1.19$& 79.9$\pm$2.6     &  79.8$\pm$2.8       &  79.8$\pm$0.1        &    $<$2.7     &   60 \\
769  &$07:00:23.4$  & $-20:32:59.2$   &  $13.20$   &  $0.73$ &$1.20$& 81.6$\pm$1.4     &  81.5$\pm$1.5       &  81.5$\pm$0.0        &    $<$2.7     &  110 \\
784  &$07:00:24.1$  & $-20:35:45.1$   &  $14.20$   &  $0.79$ &$1.26$& 99.0$\pm$2.2     &  98.8$\pm$1.3       &  98.9$\pm$0.1        &  5.6$\pm$0.9  &   55 \\
806$^a$&$07:00:24.8$  & $-20:25:54.1$   &  $13.01$   &  $0.51$ &$1.12$&100.0$\pm$2.1     &  99.0$\pm$3.8       &  99.5$\pm$0.6        &      ---      &   95 \\
1110 &$07:00:36.1$  & $-20:35:47.1$   &  $13.60$   &  $0.74$ &$1.27$& 84.0$\pm$2.4     &  84.7$\pm$2.4       &  84.3$\pm$0.5        &  6.1$\pm$0.5  &  100 \\
1118 &$07:00:36.5$  & $-20:38:57.4$   &  $13.74$   &  $0.79$ &$1.32$& 82.2$\pm$2.3     &  81.1$\pm$2.9       &  81.7$\pm$0.7        &    $<$2.7     &   74 \\
1349 &$07:00:46.3$  & $-20:28:55.7$   &  $14.16$   &  $0.83$ &$1.32$& 77.2$\pm$3.8     &  76.5$\pm$3.3       &  76.8$\pm$0.5        &  5.1$\pm$0.4  &   44 \\
1534 &$07:00:54.5$  & $-20:24:30.0$   &  $13.94$   &  $0.78$ &$1.31$& 93.8$\pm$3.6     &  91.8$\pm$4.6       &  92.8$\pm$1.3        &      ---      &   50 \\  
1616 &$07:00:58.1$  & $-20:33:24.6$   &  $13.68$   &  $0.67$ &$1.11$& 42.7$\pm$2.4     &  44.8$\pm$2.9       &  43.7$\pm$1.3        &  4.2$\pm$0.4  &  100 \\
\tableline
\tableline
\\
\end{tabular}
\par \textbf{Notes.} The columns inform, from left to right: star identification, right ascension, declination, V and b-y from this paper, and V-I from Carraro \& Patat (1995), 
two epoch heliocentric radial velocities (RV$_{1}$ and RV$_{2}$) and their mean values ($\langle$RV$\rangle$), projected rotational velocities (vsini) and spectral signal-to-noise at 6000 \AA{}.
(a) The classical Cepheid XZ CMa.
\end{table*}
\end{landscape}

\clearpage

\begin{table}
\caption{Observed Fe\,{\sc i} and Fe\,{\sc ii} lines.}
\scriptsize
\begin{tabular}{ccccccccc}\\\tableline\tableline
\label{tabelFea}
        &            &         &          & \multicolumn{5}{c}{Equivalent Widths (m\AA)}
\\\tableline
        &            &         &          & \multicolumn{5}{c}{Star} \\
\cline{5 - 9}
Element &  $\lambda$\,(\AA) & $\chi$(eV) & log $gf$  & 769 & 806 & 1110 & 1118 & 395 \\
\tableline
Fe\,{\sc i} &   5159.06  &  4.28  &$-$0.650 &  --- & --- & ---&  83&  110\\
 &   5162.27  &  4.18  &   0.079 &  158 & 148 & ---& ---&  153\\
 &   5198.71  &  2.22  &$-$2.140 &  132 & 140 & 107& ---&  125\\
 &   5242.49  &  3.63  &$-$0.970 &  105 & --- & ---& ---&  111\\
 &   5250.21  &  0.12  &$-$4.920 &  --- & --- & ---& ---&  136\\
 &   5288.52  &  3.69  &$-$1.510 &  --- &  37 &  63&  79&   78\\
 &   5307.36  &  1.61  &$-$2.970 &  --- &  94 & 111& 125&  133\\
 &   5315.05  &  4.37  &$-$1.400 &  --- & --- &  59& ---&   51\\ 
 &   5321.11  &  4.43  &$-$1.190 &   60 & --- & ---& ---&  ---\\
 &   5322.04  &  2.28  &$-$2.840 &  --- &  31 &  78& ---&  103\\
 &   5364.87  &  4.45  &   0.230 &  136 & --- & 126& 150&  143\\
 &   5367.47  &  4.42  &   0.439 &  141 & --- & 131& 135&  142\\
 &   5373.71  &  4.47  &$-$0.710 &   85 & --- &  70&  77&  ---\\
 &   5410.91  &  4.47  &   0.400 &  --- & --- & 140& ---&  ---\\
 &   5417.03  &  4.42  &$-$1.530 &   53 & --- &  33&  39&   61\\
 &   5441.34  &  4.31  &$-$1.580 &   47 &  18 &  46&  58&   65\\
 &   5522.45  &  4.21  &$-$1.400 &  67  & --- &  47&  58&   60\\
 &   5531.98  &  4.91  &$-$1.460 &  37  & --- &  31&  29&  ---\\
 &   5554.90  &  4.55  &$-$0.380 &  --- &  73 & ---& ---&  ---\\
 &   5560.21  &  4.43  &$-$1.040 &  70  & --- &  63&  64&   72\\
 &   5567.39  &  2.61  &$-$2.560 &  --- & --- &  76& ---&  110\\
 &   5576.09  &  3.43  &$-$0.850 &  135 & --- & ---& ---&  ---\\
 &   5633.95  &  4.99  &$-$0.120 &  96  &  68 & ---&  81&  ---\\
 &   5635.82  &  4.26  &$-$1.740 &  58  & --- & ---&  47&   56\\
 &   5638.26  &  4.22  &$-$0.720 &  --- & --- & ---& ---&   96\\
 &   5691.50  &  4.30  &$-$1.370 &  --- &  23 &  66&  54&   79\\
 &   5705.47  &  4.30  &$-$1.360 &  64  & --- &  68&  57&  ---\\
 &   5731.76  &  4.26  &$-$1.150 &  --- & --- & ---& ---&   76\\
 &   5806.73  &  4.61  &$-$0.900 &  73  & --- & ---&  65&   62\\
 &   5852.22  &  4.55  &$-$1.180 &  62  & --- &  48& ---&   60\\
 &   5883.82  &  3.96  &$-$1.210 &  90  & --- &  69& ---&  ---\\
 &   5934.65  &  3.93  &$-$1.020 &  97  & --- & ---& ---&   92\\
 &   6020.17  &  4.61  &$-$0.210 &  --- & 101 & ---& ---&  ---\\
 &   6024.06  &  4.55  &$-$0.060 &  --- & --- & 109& 119&  126\\
 &   6027.05  &  4.08  &$-$1.090 &  103 &  44 &  80&  87&   95\\
 &   6056.01  &  4.73  &$-$0.400 &  93  & --- &  78&  81&   86\\
 &   6065.48  &  2.61  &$-$1.530 &  157 & --- & 125& 154&  156\\
\tableline
\end{tabular}
\end{table}
 
\begin{table}
\noindent
\scriptsize
\begin{tabular}{ccccccccc}
\multicolumn{9}{c}{Table 5, continued.}\\
\tableline\tableline
        &            &         &          & \multicolumn{5}{c}{Equivalent Widths (m\AA)}
\\\tableline
        &            &         &          & \multicolumn{5}{c}{Star} \\
\cline{5 - 9}
Element &  $\lambda$\,(\AA) & $\chi$(eV) & log $gf$  & 769 & 806 & 1110 & 1118 & 395 \\
\tableline
Fe\,{\sc i}  &   6079.01  &  4.65  &$-$0.970 &  --- & --- & ---&  71&  ---\\
 &   6096.66  &  3.98  &$-$1.780 &  56  & --- &  50&  49&   62\\
 &   6120.25  &  0.91  &$-$5.950 & ---  & --- &  24& ---&  ---\\
 &   6151.62  &  2.18  &$-$3.290 &  92  & --- &  75&  90&   92\\
 &   6157.73  &  4.08  &$-$1.110 &  98  &  36 &  89&  94&   89\\
 &   6165.36  &  4.14  &$-$1.470 &  59  & --- &  59&  73&   59\\
 &   6170.51  &  4.79  &$-$0.380 & ---  &  36 & ---& ---&  ---\\
 &   6173.34  &  2.22  &$-$2.880 & 107  &  39 &  92& 112&  122\\
 &   6187.99  &  3.94  &$-$1.570 &  73  & --- &  60&  72&   80\\
 &   6200.31  &  2.60  &$-$2.440 &  118 & --- & ---& 107&  103\\
 &   6213.43  &  2.22  &$-$2.480 &  123 & --- & ---& ---&  ---\\
 &   6265.13  &  2.18  &$-$2.550 &  128 &  95 & ---& 132&  135\\
 &   6311.50  &  2.83  &$-$3.230 & ---  & --- & ---&  57&  ---\\
 &   6322.69  &  2.59  &$-$2.430 & ---  & --- &  88& ---&  106\\
 &   6380.74  &  4.19  &$-$1.320 &  87  & --- &  65&  86&   70\\
 &   6392.54  &  2.28  &$-$4.030 & ---  & --- &  29& ---&  ---\\
 &   6411.65  &  3.65  &$-$0.660 &  142 & 150 & 125& 138&  ---\\
 &   6421.35  &  2.28  &$-$2.010 &  --- & 152 & 135& 153&  150\\
 &   6430.85  &  2.18  &$-$2.010 &  155 & --- & 126& 155&  159\\
 &   6436.41  &  4.19  &$-$2.460 &  25  & --- & ---& ---&  ---\\
 &   6469.19  &  4.83  &$-$0.620 &  85  & 29  &  74&  79&   71\\
 &   6593.87  &  2.44  &$-$2.420 & 124  & --- & ---& 117&  129\\
 &   6597.56  &  4.79  &$-$0.920 &  71  & --- & ---&  55&   57\\
 &   6608.03  &  2.28  &$-$4.030 &  48  & --- &  28&  39&   37\\
 &   6646.93  &  2.61  &$-$3.990 &  45  & --- & ---& ---&   30\\
 &   6653.85  &  4.14  &$-$2.520 &  26  & --- & ---& ---&  ---\\
 &   6703.57  &  2.76  &$-$3.160 &  81  & --- & ---&  64&   59\\
 &   6739.52  &  1.56  &$-$4.950 &   37 & --- & ---&  41&  ---\\
 &   6750.15  &  2.42  &$-$2.620 &  106 & --- &  95& ---&  102\\
 &   6752.71  &  4.64  &$-$1.200 &  69  & --- &  48&  59&   59\\
 &   6806.85  &  2.73  &$-$3.210 &  63  & --- &  63& ---&   64\\
 &   6810.26  &  4.61  &$-$0.990 &  58  & --- & ---&  73&   74\\
 &   6820.37  &  4.64  &$-$1.170 &  62  & --- & ---&  58&   72\\
 &   6851.64  &  1.61  &$-$5.320 & ---  & --- & ---& ---&   24\\
 &   6858.15  &  4.61  &$-$0.930 &  77  & --- &  68&  63&  ---\\
 &   7130.92  &  4.22  &$-$0.700 &  115 &  85 & ---& ---&  103\\
 &   7132.99  &  4.08  &$-$1.610 &   63 &  25 &  49& ---&   65\\
Fe\,{\sc ii} &  5132.66  &  2.81  &  -4.000 & --- & --- &  44& ---&  ---\\
  &  5425.25  &  3.20  &$-$3.210 & --- & --- & ---&  63&   67\\
\tableline
\end{tabular}
\end{table}
 
\begin{table}
\noindent
\scriptsize
\begin{tabular}{ccccccccc}
\multicolumn{9}{c}{Table 5, continued.}\\
\tableline\tableline
        &            &         &          & \multicolumn{5}{c}{Equivalent Widths (m\AA)}
\\\tableline
        &            &         &          & \multicolumn{5}{c}{Star} \\
\cline{5 - 9}
Element &  $\lambda$\,(\AA) & $\chi$(eV) & log $gf$  & 769 & 806 & 1110 & 1118 & 395 \\
\tableline
Fe\,{\sc ii}  &  5991.37  &  3.15  &$-$3.560 & --- & --- &  43& ---&  ---\\
  &  6084.10  &  3.20  &$-$3.800 & --- & --- &  35&  45&  ---\\
  &  6149.25  &  3.89  &$-$2.720 & --- & --- &  43& ---&   60\\
  &  6247.55  &  3.89  &$-$2.340 &  81 & 139 & ---&  74&   82\\
  &  6416.92  &  3.89  &$-$2.680 &  54 &  96 &  49& ---&   52\\
  &  6432.68  &  2.89  &$-$3.580 &  61 & --- &  50& ---&   63\\
\tableline
\end{tabular}
\end{table}

\clearpage

\begin{table}
\caption{Observed Fe\,{\sc i} and Fe\,{\sc ii} lines.}
\scriptsize
\begin{tabular}{ccccccccc}\\\tableline\tableline
\label{tabelFeb}
        &            &         &          & \multicolumn{5}{c}{Equivalent Widths (m\AA)}
\\\tableline
        &            &         &          & \multicolumn{5}{c}{Star} \\
\cline{5 - 9}
Element &  $\lambda$\,(\AA) & $\chi$(eV) & log $gf$  & 1534 & 1616 & 1349 & 784 & 663 \\
\tableline
Fe\,{\sc i} &   5159.06  &  4.28  &$-$0.650 &  --- &  98 & 106& 109&  ---\\
 &   5253.03  &  2.28  &$-$3.790 &  --- & --- & ---& ---&   91\\
 &   5288.52  &  3.69  &$-$1.510 &  --- & --- & ---&  93&  116\\
 &   5307.36  &  1.61  &$-$2.970 &  --- & 109 & 147& 148&  ---\\
 &   5315.05  &  4.37  &$-$1.400 &  --- & --- & ---& ---&   70\\ 
 &   5321.11  &  4.43  &$-$1.190 &  --- & --- & ---& ---&   98\\
 &   5322.04  &  2.28  &$-$2.840 &  121 & --- & ---& ---&  ---\\
 &   5364.87  &  4.45  &   0.230 &  158 & --- & 148& 154&  ---\\
 &   5367.47  &  4.42  &   0.439 &  157 & --- & ---& ---&  ---\\
 &   5373.71  &  4.47  &$-$0.710 &  116 & --- & 113&  89&  116\\
 &   5417.03  &  4.42  &$-$1.530 &   53 &  52 &  60&  61&   80\\
 &   5441.34  &  4.31  &$-$1.580 &   44 &  40 &  57& ---&  ---\\
 &   5522.45  &  4.21  &$-$1.400 &  --- & --- &  68&  71&   92\\
 &   5554.90  &  4.55  &$-$0.380 &  --- & 103 & 122& ---&  ---\\
 &   5560.21  &  4.43  &$-$1.040 &  --- & --- & 104& ---&  ---\\
 &   5567.39  &  2.61  &$-$2.560 &  122 & --- & ---& ---&  155\\
 &   5576.09  &  3.43  &$-$0.850 &  --- & --- & ---& 154&  ---\\
 &   5624.02  &  4.39  &$-$1.330 &  --- & --- & ---&  76&  103\\
 &   5633.95  &  4.99  &$-$0.120 &  --- &  85 & 103& ---&  ---\\
 &   5635.82  &  4.26  &$-$1.740 &  --- & --- & ---& ---&   58\\
 &   5638.26  &  4.22  &$-$0.720 &  --- &  84 & ---& 106&  ---\\
 &   5691.50  &  4.30  &$-$1.370 &  85  &  51 & ---&  74&  104\\
 &   5705.47  &  4.30  &$-$1.360 &  --- &  46 & ---&  59&   93\\
 &   5717.83  &  4.28  &$-$0.979 &  116 & --- & 122& ---&  ---\\
 &   5806.73  &  4.61  &$-$0.900 &  --- & --- & ---&  65&   99\\
 &   5814.81  &  4.28  &$-$1.820 &  --- & --- & ---& ---&   62\\
 &   5852.22  &  4.55  &$-$1.180 &  --- &  63 &  78&  70&   95\\
 &   5883.82  &  3.96  &$-$1.210 &  94  & --- & ---& 101&  ---\\
 &   5916.25  &  2.45  &$-$2.990 & 104  & --- & ---& ---&  ---\\
 &   5934.65  &  3.93  &$-$1.020 & 120  & --- & 142& 117&  140\\
 &   6024.06  &  4.55  &$-$0.060 &  --- & 102 & ---& ---&  ---\\
 &   6027.05  &  4.08  &$-$1.090 &  --- &  72 & ---& ---&  ---\\
 &   6056.01  &  4.73  &$-$0.400 &  --- &  91 & ---& ---&  123\\
 &   6065.48  &  2.61  &$-$1.530 &  --- & 151 & ---& ---&  ---\\
 &   6096.66  &  3.98  &$-$1.780 &  --- &  62 &  95&  71&   83\\
 &   6151.62  &  2.18  &$-$3.290 &  --- &  79 & ---& ---&  ---\\
 &   6157.73  &  4.08  &$-$1.110 &  120 &  95 & ---& 102&  120\\
\tableline
\end{tabular}
\end{table}
 
\begin{table}
\noindent
\scriptsize
\begin{tabular}{ccccccccc}
\multicolumn{9}{c}{Table 6, continued.}\\
\tableline\tableline
        &            &         &          & \multicolumn{5}{c}{Equivalent Widths (m\AA)}
\\\tableline
        &            &         &          & \multicolumn{5}{c}{Star} \\
\cline{5 - 9}
Element &  $\lambda$\,(\AA) & $\chi$(eV) & log $gf$  & 1534 & 1616 & 1349 & 784 & 663 \\
\tableline
Fe\,{\sc i}  &   6165.36  &  4.14  &$-$1.470 &  --- &  56 & ---&  75&   97\\
 &   6173.34  &  2.22  &$-$2.880 & ---  &  98 & ---& ---&  145\\
 &   6187.99  &  3.94  &$-$1.570 & ---  & --- & ---&  81&  102\\
 &   6200.31  &  2.60  &$-$2.440 &  --- & 111 & 145& 126&  ---\\
 &   6265.13  &  2.18  &$-$2.550 &  155 & --- & ---& 143&  ---\\
 &   6380.74  &  4.19  &$-$1.320 &  --- &  75 & ---& ---&  119\\
 &   6392.54  &  2.28  &$-$4.030 & ---  & --- & ---&  65&   84\\
 &   6411.65  &  3.65  &$-$0.660 &  --- & 145 & ---& 157&  ---\\
 &   6421.35  &  2.28  &$-$2.010 &  --- & 133 & ---& ---&  ---\\
 &   6430.85  &  2.18  &$-$2.010 &  --- & 148 & ---& ---&  ---\\
 &   6436.41  &  4.19  &$-$2.460 &  --- & --- & ---&  21&   42\\
 &   6469.19  &  4.83  &$-$0.620 &  70  &  82 & 102&  99&  117\\
 &   6551.68  &  0.99  &$-$5.790 & ---  & --- &  56&  41&   65\\
 &   6591.31  &  4.59  &$-$2.070 & ---  & --- & ---&  23&  ---\\
 &   6593.87  &  2.44  &$-$2.420 & 137  & --- & ---& ---&  ---\\
 &   6597.56  &  4.79  &$-$0.920 &  52  &  40 & ---&  62&   73\\
 &   6608.03  &  2.28  &$-$4.030 & ---  &  40 & ---&  66&   76\\
 &   6609.11  &  2.56  &$-$2.690 & ---  & --- & 140& ---&  ---\\
 &   6646.93  &  2.61  &$-$3.990 &  41  & --- & ---& ---&   63\\
 &   6703.57  &  2.76  &$-$3.160 & ---  & --- &  66&  79&  ---\\
 &   6739.52  &  1.56  &$-$4.950 &  --- & --- & ---&  70&   48\\
 &   6750.15  &  2.42  &$-$2.620 &  --- & --- & 150& ---&  ---\\
 &   6752.71  &  4.64  &$-$1.200 &  --- &  38 &  86&  73&  101\\
 &   6806.85  &  2.73  &$-$3.210 &  --- &  59 &  90&  87&  104\\
 &   6810.26  &  4.61  &$-$0.990 &  --- & --- &  98&  91&   94\\
 &   6820.37  &  4.64  &$-$1.170 &  --- &  58 &  65& ---&   83\\
 &   6858.15  &  4.61  &$-$0.930 &  --- &  65 & ---& ---&  103\\
 &   7130.92  &  4.22  &$-$0.700 &  --- & --- & ---& 130&  ---\\
 &   7132.99  &  4.08  &$-$1.610 &  --- & --- &  77&  64&   90\\
Fe\,{\sc ii} &  5425.25  &  3.20  &$-$3.210 & --- & --- & ---&  69&  ---\\
  &  5534.83  &  3.25  &$-$2.770 & --- &  87 & ---& 102&  ---\\
  &  6084.10  &  3.20  &$-$3.800 & --- &  25 & ---& ---&  ---\\
  &  6149.25  &  3.89  &$-$2.720 &  73 &  54 &  80&  63&  ---\\
  &  6247.55  &  3.89  &$-$2.340 & --- &  68 & ---&  82&   73\\
  &  6416.92  &  3.89  &$-$2.680 &  78 &  63 &  69&  59&   56\\
  &  6432.68  &  2.89  &$-$3.580 & --- & --- & ---&  68&   53\\
\tableline
\end{tabular}
\end{table}

\clearpage

\begin{table*}
\caption{Other lines studied.}
\tiny
\begin{tabular}{cccccccccc}
\label{tabellinesa}
\\\tableline\tableline
    & & & & &\multicolumn{5}{c}{Equivalent Widths (m\AA)} \\\tableline
\multicolumn{5}{c}{} & \multicolumn{5}{c}{Star}\\
\cline{6-10}
Element & $\lambda$ & $\chi$(eV) & $\log gf$ & Ref & 769 & 806 & 1110 & 1118 & 395 \\
\tableline
Na\,{\sc i} & 6154.22 & 2.10 & $-$1.51 & PS   &  59 &  17 &  63 &  72 &  55 \\
Na\,{\sc i} & 6160.75 & 2.10 & $-$1.21 & R03  &  89 & --- &  72 &  86 &  78 \\
Mg\,{\sc i} & 4730.04 & 4.34 & $-$2.39 & R03  &  81 & --- & --- & --- & --- \\ 
Mg\,{\sc i} & 5711.10 & 4.34 & $-$1.75 & R99  & 115 &  77 & 107 & --- & --- \\
Mg\,{\sc i} & 6318.71 & 5.11 & $-$1.94 & Ca07 & --- & --- &  45 &  58 &  55 \\
Mg\,{\sc i} & 6965.41 & 5.75 & $-$1.72 & MR94 & --- & --- & --- &  43 &  40 \\
Mg\,{\sc i} & 7387.70 & 5.75 & $-$0.87 & MR94 &  85 & --- & 105 & --- &  80 \\
Mg\,{\sc i} & 8717.83 & 5.91 & $-$0.97 & WSM  &  62 & --- & --- &  83 & --- \\
Mg\,{\sc i} & 8736.04 & 5.94 & $-$0.34 & WSM  & --- & --- & --- & 128 & --- \\
Si\,{\sc i} & 5793.08 & 4.93 & $-$2.06 &  R03 &  52 &  17 & --- & 61  &  63 \\
Si\,{\sc i} & 6125.03 & 5.61 & $-$1.54 &  E93 &  44 &  24 &  29 & 39  &  42 \\
Si\,{\sc i} & 6131.58 & 5.62 & $-$1.68 &  E93 & --- &  14 & --- & --- & --- \\
Si\,{\sc i} & 6155.14 & 5.62 & $-$0.77 &  E93 &  91 &  77 & --- & 87  &  89 \\
Si\,{\sc i} & 7800.00 & 6.18 & $-$0.72 &  E93 &  69 & --- &  58 & --- &  68 \\
Si\,{\sc i} & 8728.01 & 6.18 & $-$0.36 &  E93 &  89 & --- & --- & --- & --- \\
Ca\,{\sc i} & 6102.73 & 1.88 & $-$0.79 & D2002& --- & --- & --- & 149 & 141 \\
Ca\,{\sc i} & 6161.30 & 2.52 & $-$1.27 &  E93 &  93 & --- &  78 &  92 &  97 \\
Ca\,{\sc i} & 6166.44 & 2.52 & $-$1.14 &  R03 &  83 &  43 &  88 &  99 &  86 \\
Ca\,{\sc i} & 6169.04 & 2.52 & $-$0.80 &  R03 & 114 & --- &  93 & 113 & 112 \\
Ca\,{\sc i} & 6169.56 & 2.53 & $-$0.48 & DS91 & 126 & --- & 121 & 125 & 127 \\
Ca\,{\sc i} & 6455.60 & 2.51 & $-$1.29 &  R03 &  79 & --- &  71 & --- &  89 \\
Ca\,{\sc i} & 6471.66 & 2.51 & $-$0.69 &  S86 & 132 &  98 & --- & --- & 125 \\
Ca\,{\sc i} & 6493.79 & 2.52 & $-$0.11 & DS91 & 152 & --- & --- & --- & --- \\
Ti\,{\sc i} & 4534.78 & 0.84 &    0.280 &D2002& 149 & --- & --- & --- & --- \\
Ti\,{\sc i} & 4758.12 & 2.25 &    0.420 & MFK &  80 & --- & --- & --- & --- \\
Ti\,{\sc i} & 4759.28 & 2.25 &    0.514 & MFK &  81 & --- & --- & --- & --- \\
Ti\,{\sc i} & 4820.41 & 1.50 & $-$0.439 & MFK &  83 & --- & --- & --- &  83 \\
Ti\,{\sc i} & 4999.51 & 0.83 &    0.250 & MFK & --- & --- & 119 & --- & --- \\
Ti\,{\sc i} & 5009.66 & 0.02 & $-$2.259 & MFK & --- & --- & --- &  71 & --- \\
Ti\,{\sc i} & 5022.87 & 0.83 & $-$0.434 & MFK & --- & --- & --- & --- & 109 \\
Ti\,{\sc i} & 5039.96 & 0.02 & $-$1.130 & MFK & 118 & --- &  96 & --- & --- \\
Ti\,{\sc i} & 5043.59 & 0.84 & $-$1.733 & MFK &  63 & --- & --- & --- & --- \\
Ti\,{\sc i} & 5087.06 & 1.43 & $-$0.840 & E93 &  60 & --- & --- &  56 &  56 \\
Ti\,{\sc i} & 5113.45 & 1.44 & $-$0.880 & E93 & --- & --- & --- &  59 & --- \\
Ti\,{\sc i} & 5145.47 & 1.46 & $-$0.574 & MFK &  82 & --- &  62 & --- &  67 \\
Ti\,{\sc i} & 5147.48 & 0.00 & $-$2.012 & MFK &  82 & --- & --- &  97 &  94 \\
Ti\,{\sc i} & 5173.75 & 0.00 & $-$1.120 & MFK & --- & --- & --- & 136 & --- \\
Ti\,{\sc i} & 5210.39 & 0.05 & $-$0.883 & MFK & --- & --- & 102 & --- & --- \\
Ti\,{\sc i} & 5219.71 & 0.02 & $-$2.292 & MFK &  81 & --- & --- & --- &  79 \\
Ti\,{\sc i} & 5223.63 & 2.09 & $-$0.559 & MFK & --- & --- &  29 & --- & --- \\
Ti\,{\sc i} & 5295.78 & 1.05 & $-$1.633 & MFK & --- & --- & --- &  39 &  44 \\
Ti\,{\sc i} & 5490.16 & 1.46 & $-$0.937 & MFK &  70 & --- & --- & --- & --- \\
Ti\,{\sc i} & 5689.48 & 2.30 & $-$0.469 & MFK &  34 & --- &  35 & --- &  40 \\
Ti\,{\sc i} & 5866.46 & 1.07 & $-$0.871 & E93 & --- & --- &  77 &  82 &  93 \\
Ti\,{\sc i} & 5922.12 & 1.05 & $-$1.465 & MFK &  62 & --- & --- &  59 & --- \\
Ti\,{\sc i} & 5978.55 & 1.87 & $-$0.496 & MFK &  63 & --- &  52 &  50 & --- \\
Ti\,{\sc i} & 6091.18 & 2.27 & $-$0.370 & R03 &  43 & --- & --- &  42 & --- \\
\tableline
\end{tabular}
\end{table*}

\begin{table*}
\noindent
\tiny
\begin{tabular}{cccccccccc}
\multicolumn{10}{c}{Table 7, continued.}\\
\tableline\tableline
    & & & & &\multicolumn{5}{c}{Equivalent Widths (m\AA)} \\\tableline
\multicolumn{5}{c}{} & \multicolumn{5}{c}{Star}\\
\cline{6-10}
Element & $\lambda$ & $\chi$(eV) & $\log gf$ & Ref & 769 & 806 & 1110 & 1118 & 395 \\
\tableline
Ti\,{\sc i} & 6126.22 & 1.07 & $-$1.370 & R03 &  61 &  10 &  45 & --- &  60 \\
Ti\,{\sc i} & 6258.11 & 1.44 & $-$0.355 & MFK &  99 & --- &  81 &  81 & --- \\
Ti\,{\sc i} & 6261.11 & 1.43 & $-$0.480 & B86 &  96 & --- &  76 &  83 &  89 \\
Ti\,{\sc i} & 6554.24 & 1.44 & $-$1.219 & MFK & --- & --- & --- &  55 & --- \\
Cr\,{\sc i} & 4836.85 & 3.10 & $-$1.137 & MFK &  40 & --- & --- & --- &  32 \\
Cr\,{\sc i} & 5200.18 & 3.38 & $-$0.650 & MFK &  40 & --- & --- & --- & --- \\
Cr\,{\sc i} & 5296.70 & 0.98 & $-$1.390 & GS  & 133 &  82 & --- & 134 & 132 \\
Cr\,{\sc i} & 5304.18 & 3.46 & $-$0.692 & MFK &  35 & --- & --- & --- & --- \\
Cr\,{\sc i} & 5345.81 & 1.00 & $-$0.980 & GS  & --- & 145 & 140 & --- & --- \\
Cr\,{\sc i} & 5348.32 & 1.00 & $-$1.290 & GS  & --- &  85 & 118 & 135 & 147 \\
Cr\,{\sc i} & 5783.07 & 3.32 & $-$0.500 & MFK &  49 & --- &  47 &  60 &  62 \\
Cr\,{\sc i} & 5783.87 & 3.32 & $-$0.290 & GS  &  74 & --- &  64 &  68 &  76 \\
Cr\,{\sc i} & 5787.93 & 3.32 & $-$0.080 & GS  &  71 & --- &  69 & --- &  70 \\
Cr\,{\sc i} & 6330.09 & 0.94 & $-$2.920 & R03 & --- & --- &  58 &  61 &  70 \\
Ni\,{\sc i} & 4904.42 & 3.54 & $-$0.170 & MFK & 116 & --- & 102 & 100 & --- \\
Ni\,{\sc i} & 4935.83 & 3.94 & $-$0.360 & MFK &  82 & --- &  60 &  85 & --- \\
Ni\,{\sc i} & 4953.21 & 3.74 & $-$0.660 & MFK &  87 & --- & --- &  72 &  81 \\
Ni\,{\sc i} & 4967.52 & 3.80 & $-$1.570 & MFK &  40 & --- & --- & --- & --- \\
Ni\,{\sc i} & 5010.94 & 3.63 & $-$0.870 & MFK & --- & --- &  62 &  73 &  63 \\
Ni\,{\sc i} & 5084.11 & 3.68 & $-$0.180 & E93 & 100 & --- &  86 & --- &  95 \\
Ni\,{\sc i} & 5094.42 & 3.83 & $-$1.080 & MFK &  51 & --- &  38 & --- & --- \\
Ni\,{\sc i} & 5115.40 & 3.83 & $-$0.280 & R03 & 101 & --- & --- & --- &  90 \\
Ni\,{\sc i} & 5157.98 & 3.61 & $-$1.590 & MFK & --- & --- &  33 & --- & --- \\
Ni\,{\sc i} & 5578.73 & 1.68 & $-$2.640 & MFK &  93 & --- & --- & --- &  91 \\
Ni\,{\sc i} & 5589.37 & 3.90 & $-$1.140 & MFK &  42 & --- &  35 &  45 &  37 \\
Ni\,{\sc i} & 5593.75 & 3.90 & $-$0.840 & MFK &  56 & --- & --- &  61 &  68 \\
Ni\,{\sc i} & 5760.84 & 4.11 & $-$0.800 & MFK &  68 & --- & --- &  63 & --- \\
Ni\,{\sc i} & 5805.23 & 4.17 & $-$0.640 & MFK &  56 &  37 & --- & --- &  52 \\
Ni\,{\sc i} & 5996.74 & 4.24 & $-$1.060 & MFK &  30 & --- & --- & --- &  30 \\
Ni\,{\sc i} & 6053.69 & 4.24 & $-$1.070 & MFK & --- & --- & --- &  39 & --- \\
Ni\,{\sc i} & 6086.29 & 4.27 & $-$0.510 & MFK &  66 & --- &  43 &  66 &  66 \\
Ni\,{\sc i} & 6108.12 & 1.68 & $-$2.440 & MFK & 106 & --- &  90 & --- & --- \\
Ni\,{\sc i} & 6111.08 & 4.09 & $-$0.870 & MFK &  68 & --- &  48 &  49 & --- \\
Ni\,{\sc i} & 6128.98 & 1.68 & $-$3.320 & MFK &  66 & --- &  51 & --- & --- \\
Ni\,{\sc i} & 6176.82 & 4.09 & $-$0.264 & R03 & --- & --- & --- &  77 & --- \\
Ni\,{\sc i} & 6186.72 & 4.11 & $-$0.960 & MFK &  48 & --- &  40 &  46 & --- \\
Ni\,{\sc i} & 6204.61 & 4.09 & $-$1.150 & MFK &  52 & --- &  23 & --- &  42 \\
Ni\,{\sc i} & 6223.99 & 4.11 & $-$0.980 & MFK & --- & --- & --- & --- &  41 \\
Ni\,{\sc i} & 6230.10 & 4.11 & $-$1.260 & MFK &  40 & --- & --- &  26 &  37 \\
Ni\,{\sc i} & 6322.17 & 4.15 & $-$1.170 & MFK &  33 & --- &  31 & --- & --- \\
Ni\,{\sc i} & 6327.60 & 1.68 & $-$3.150 & MFK &  84 & --- & --- &  79 &  81 \\
Ni\,{\sc i} & 6378.26 & 4.15 & $-$0.900 & MFK &  60 & --- & --- & --- &  44 \\
Ni\,{\sc i} & 6482.81 & 1.94 & $-$2.630 & MFK & --- &  35 & --- & --- & --- \\
Ni\,{\sc i} & 6532.88 & 1.94 & $-$3.390 & MFK &  62 & --- & --- & --- & --- \\
Ni\,{\sc i} & 6586.32 & 1.95 & $-$2.810 & MFK &  74 & --- &  71 &  92 & --- \\
Ni\,{\sc i} & 6635.14 & 4.42 & $-$0.830 & MFK & --- & --- &  27 & --- & --- \\
Ni\,{\sc i} & 6643.64 & 1.68 & $-$2.030 & MFK & 142 & --- & 109 & 150 & 136 \\
Ni\,{\sc i} & 6767.78 & 1.83 & $-$2.170 & MFK & 111 &  76 & 100 & 118 & 113 \\
\tableline
\end{tabular}
\end{table*}

\begin{table*}
\noindent
\tiny
\begin{tabular}{cccccccccc}
\multicolumn{10}{c}{Table 7, continued.}\\
\tableline\tableline
    & & & & &\multicolumn{5}{c}{Equivalent Widths (m\AA)} \\\tableline
\multicolumn{5}{c}{} & \multicolumn{5}{c}{Star}\\
\cline{6-10}
Element & $\lambda$ & $\chi$(eV) & $\log gf$ & Ref & 769 & 806 & 1110 & 1118 & 395 \\
\tableline
Ni\,{\sc i} & 6772.32 & 3.66 & $-$0.970 & R03 & --- & --- &  55 &  79 & --- \\
Ni\,{\sc i} & 6842.04 & 3.66 & $-$1.477 & E93 &  49 & --- & --- &  42 &  53 \\
Ni\,{\sc i} & 7788.93 & 1.95 & $-$1.990 & E93 & --- & --- & --- & --- & 122 \\
\tableline
\\
\end{tabular}
\par References: B86: Blackwell D.E. et al. (1986); Ca07: Carretta et al. (2007);
\par D2002; Depagne et al. (2002); DS91: Drake \& Smith (1991);
\par E93: Edvardsson et al. (1993); GS: Gratton \& Sneden (1988);
\par MFK: Martin et al., (2002); MR94: Mcwilliam \& Rich (1994);
\par PS: Preston \& Sneden (2001); R03: Reddy et al. (2003);
\par R99: Reddy et al. (1999); WSM: Wiese, Smith \& Miles (1969).

\end{table*}

\clearpage

\begin{table*}
\caption{Other lines studied.}
\tiny
\begin{tabular}{cccccccccc}
\label{tabellinesb}
\\\tableline\tableline
    & & & & &\multicolumn{5}{c}{Equivalent Widths (m\AA)} \\\tableline
\multicolumn{5}{c}{} & \multicolumn{5}{c}{Star}\\
\cline{6-10}
Element & $\lambda$ & $\chi$(eV) & $\log gf$ & Ref & 1534 & 1616 & 1349 & 784 & 663 \\
\tableline
Na\,{\sc i} & 6154.22 & 2.10 & $-$1.51 & PS   &  47 &  68 & --- &  68 & 105 \\
Na\,{\sc i} & 6160.75 & 2.10 & $-$1.21 & R03  & --- & --- & --- &  86 & 122 \\
Mg\,{\sc i} & 4730.04 & 4.34 & $-$2.39 & R03  &  88 & --- & --- &  79 & --- \\ 
Mg\,{\sc i} & 5711.10 & 4.34 & $-$1.75 & R99  & 140 & 124 & 124 & 138 & 152 \\
Mg\,{\sc i} & 6318.71 & 5.11 & $-$1.94 & Ca07 & --- &  58 & --- &  74 &  91 \\
Mg\,{\sc i} & 7387.70 & 5.75 & $-$0.87 & MR94 & --- & 102 &  98 & 113 & 120 \\
Mg\,{\sc i} & 8736.04 & 5.94 & $-$0.34 & WSM  & 133 & --- & --- & --- & --- \\
Al\,{\sc i} & 6698.67 & 3.14 & $-$1.63 & R03  & --- & --- &  58 & --- & --- \\
Al\,{\sc i} & 7835.32 & 4.04 & $-$0.58 & R03  & --- &  39 & --- &  69 &  78 \\
Al\,{\sc i} & 7836.13 & 4.02 & $-$0.40 & R03  & --- &  42 & --- &  80 & 102 \\
Al\,{\sc i} & 8772.88 & 4.02 & $-$0.25 & R03  &  99 & --- &  99 & --- & --- \\
Al\,{\sc i} & 8773.91 & 4.02 & $-$0.07 & R03  & 114 & --- & --- & --- & 140 \\
Si\,{\sc i} & 5793.08 & 4.93 & $-$2.06 &  R03 &  53 &  32 &  48 & 76  &  86 \\
Si\,{\sc i} & 6125.03 & 5.61 & $-$1.54 &  E93 &  31 &  26 & --- & 50  & --- \\
Si\,{\sc i} & 6145.02 & 5.61 & $-$1.43 &  E93 & --- & --- & --- & 59  &  64 \\
Si\,{\sc i} & 6155.14 & 5.62 & $-$0.77 &  E93 &  79 &  81 & --- & 96  & --- \\
Si\,{\sc i} & 7800.00 & 6.18 & $-$0.72 &  E93 &  40 & --- & --- & 75  & --- \\
Si\,{\sc i} & 8728.01 & 6.18 & $-$0.36 &  E93 &  75 & --- & --- & --- & 122 \\
Si\,{\sc i} & 8742.45 & 5.87 & $-$0.51 &  E93 & --- & --- & 106 & --- & 119 \\
Ca\,{\sc i} & 6102.73 & 1.88 & $-$0.79 & D2002& --- & --- & --- & 160 & --- \\
Ca\,{\sc i} & 6161.30 & 2.52 & $-$1.27 &  E93 &  82 & --- & --- & 101 & 137 \\
Ca\,{\sc i} & 6166.44 & 2.52 & $-$1.14 &  R03 &  95 &  80 & 118 & 101 & 128 \\
Ca\,{\sc i} & 6169.04 & 2.52 & $-$0.80 &  R03 & --- & --- & --- & 114 & --- \\
Ca\,{\sc i} & 6169.56 & 2.53 & $-$0.48 & DS91 & --- & --- & --- & 142 & --- \\
Ca\,{\sc i} & 6455.60 & 2.51 & $-$1.29 &  R03 & --- &  80 & 101 &  90 & --- \\
Ca\,{\sc i} & 6471.66 & 2.51 & $-$0.69 &  S86 & 122 & 114 & --- & --- & --- \\
Ca\,{\sc i} & 6493.79 & 2.52 & $-$0.11 & DS91 & 144 & --- & --- & --- & --- \\
Ti\,{\sc i} & 4759.28 & 2.25 &    0.514 & MFK & --- &  86 & --- &  88 & --- \\
Ti\,{\sc i} & 4820.41 & 1.50 & $-$0.439 & MFK & --- &  65 & --- &  92 & --- \\
Ti\,{\sc i} & 5009.66 & 0.02 & $-$2.259 & MFK & --- & --- & 112 &  88 & --- \\
Ti\,{\sc i} & 5039.96 & 0.02 & $-$1.130 & MFK & --- & 123 & --- & --- & --- \\
Ti\,{\sc i} & 5043.59 & 0.84 & $-$1.733 & MFK & --- & --- &  73 & --- & 111 \\
Ti\,{\sc i} & 5062.10 & 2.16 & $-$0.464 & MFK &  35 & --- & --- & --- &  64 \\
Ti\,{\sc i} & 5113.45 & 1.44 & $-$0.880 & E93 & --- & --- & --- & --- & 106 \\
Ti\,{\sc i} & 5147.48 & 0.00 & $-$2.012 & MFK & --- &  89 & --- & --- & --- \\
Ti\,{\sc i} & 5210.39 & 0.05 & $-$0.883 & MFK & --- & --- & --- & 156 & --- \\
Ti\,{\sc i} & 5219.71 & 0.02 & $-$2.292 & MFK & --- & --- & 117 & --- & --- \\
Ti\,{\sc i} & 5295.78 & 1.05 & $-$1.633 & MFK & --- &  36 & --- &  56 &  64 \\
Ti\,{\sc i} & 5689.48 & 2.30 & $-$0.469 & MFK & --- & --- & --- &  38 &  67 \\
Ti\,{\sc i} & 5866.46 & 1.07 & $-$0.871 & E93 & --- &  91 & --- & --- & 155 \\
Ti\,{\sc i} & 5922.12 & 1.05 & $-$1.465 & MFK & --- & --- & --- & --- & 101 \\
Ti\,{\sc i} & 5978.55 & 1.87 & $-$0.496 & MFK & --- &  43 & --- &  76 & --- \\
Ti\,{\sc i} & 6126.22 & 1.07 & $-$1.370 & R03 & --- &  44 & --- &  71 & --- \\
Ti\,{\sc i} & 6258.11 & 1.44 & $-$0.355 & MFK & --- &  96 & --- & --- & 149 \\
Ti\,{\sc i} & 6261.11 & 1.43 & $-$0.480 & B86 &  95 &  86 & 124 & 105 & --- \\
Ti\,{\sc i} & 6554.24 & 1.44 & $-$1.219 & MFK & --- & --- & --- &  51 &  84 \\
\tableline
\end{tabular}
\end{table*}

\begin{table}[!h]
\noindent
\tiny
\begin{tabular}{cccccccccc}
\multicolumn{10}{c}{Table 8, continued.}\\
\tableline\tableline
    & & & & &\multicolumn{5}{c}{Equivalent Widths (m\AA)} \\\tableline
\multicolumn{5}{c}{} & \multicolumn{5}{c}{Star}\\
\cline{6-10}
Element & $\lambda$ & $\chi$(eV) & $\log gf$ & Ref & 1534 & 1616 & 1349 & 784 & 663 \\
\tableline
Cr\,{\sc i} & 4836.85 & 3.10 & $-$1.137 & MFK & --- & --- &  49 & --- & --- \\
Cr\,{\sc i} & 5193.50 & 3.42 & $-$0.720 & MFK & --- & --- & --- &  23 & --- \\
Cr\,{\sc i} & 5196.45 & 3.45 & $-$0.270 & MFK & --- & --- &  92 & --- & --- \\
Cr\,{\sc i} & 5214.13 & 3.37 & $-$0.740 & MFK & --- & --- & --- &  32 & --- \\
Cr\,{\sc i} & 5296.70 & 0.98 & $-$1.390 & GS  & --- & 129 & --- & --- & --- \\
Cr\,{\sc i} & 5348.32 & 1.00 & $-$1.290 & GS  & --- & 126 & --- & 154 & --- \\
Cr\,{\sc i} & 5702.32 & 3.45 & $-$0.666 & MFK &  21 & --- &  63 & --- & --- \\
Cr\,{\sc i} & 5783.07 & 3.32 & $-$0.500 & MFK & --- & --- & --- &  54 &  75 \\
Cr\,{\sc i} & 5783.87 & 3.32 & $-$0.290 & GS  &  49 &  60 &  95 &  57 &  90 \\
Cr\,{\sc i} & 5787.92 & 3.32 & $-$0.080 & GS  &  51 & --- & 100 & --- & 103 \\
Cr\,{\sc i} & 6330.09 & 0.94 & $-$2.920 & R03 & --- &  67 & --- & --- & --- \\
Ni\,{\sc i} & 4904.42 & 3.54 & $-$0.170 & MFK & 117 & --- & 139 & --- & --- \\
Ni\,{\sc i} & 4935.83 & 3.94 & $-$0.360 & MFK & --- & --- &  99 &  78 & --- \\
Ni\,{\sc i} & 4953.21 & 3.74 & $-$0.660 & MFK &  61 &  65 & --- & --- & 105 \\
Ni\,{\sc i} & 4967.52 & 3.80 & $-$1.570 & MFK & --- & --- & --- &  35 & --- \\
Ni\,{\sc i} & 5010.94 & 3.63 & $-$0.870 & MFK & --- & --- &  78 &  68 & --- \\
Ni\,{\sc i} & 5048.85 & 3.85 & $-$0.370 & MFK &  94 & --- & --- & --- & --- \\
Ni\,{\sc i} & 5084.11 & 3.68 & $-$0.180 & E93 & --- &  94 & --- & --- & --- \\
Ni\,{\sc i} & 5094.42 & 3.83 & $-$1.080 & MFK &  51 &  53 & --- & --- & --- \\
Ni\,{\sc i} & 5115.40 & 3.83 & $-$0.280 & R03 & --- &  95 & --- & --- & --- \\
Ni\,{\sc i} & 5157.98 & 3.61 & $-$1.590 & MFK & --- & --- &  44 & --- & --- \\
Ni\,{\sc i} & 5578.73 & 1.68 & $-$2.640 & MFK & --- & --- & --- & 105 & --- \\
Ni\,{\sc i} & 5589.37 & 3.90 & $-$1.140 & MFK & --- & --- &  46 & --- & --- \\
Ni\,{\sc i} & 5593.75 & 3.90 & $-$0.840 & MFK & --- &  51 &  68 & --- &  84 \\
Ni\,{\sc i} & 5643.09 & 4.17 & $-$1.250 & MFK & --- & --- &  27 & --- & --- \\
Ni\,{\sc i} & 5748.36 & 1.68 & $-$3.260 & MFK & --- &  64 & --- & --- & 113 \\
Ni\,{\sc i} & 5760.84 & 4.11 & $-$0.800 & MFK & --- &  59 &  61 & --- & --- \\
Ni\,{\sc i} & 5805.23 & 4.17 & $-$0.640 & MFK & --- & --- & --- & --- &  59 \\
Ni\,{\sc i} & 5996.74 & 4.24 & $-$1.060 & MFK & --- & --- & --- & --- &  57 \\
Ni\,{\sc i} & 6086.29 & 4.27 & $-$0.510 & MFK & --- & --- & --- &  77 & --- \\
Ni\,{\sc i} & 6108.12 & 1.68 & $-$2.440 & MFK & 124 &  90 & 140 & --- & --- \\
Ni\,{\sc i} & 6111.08 & 4.09 & $-$0.870 & MFK & --- & --- & --- & --- &  75 \\
Ni\,{\sc i} & 6128.98 & 1.68 & $-$3.320 & MFK & --- &  48 & --- &  69 & 105 \\
Ni\,{\sc i} & 6176.82 & 4.09 & $-$0.264 & R03 & --- & --- & --- &  86 & 114 \\
Ni\,{\sc i} & 6177.25 & 1.83 & $-$3.510 & MFK & --- & --- & --- &  45 &  67 \\
Ni\,{\sc i} & 6186.72 & 4.11 & $-$0.960 & MFK &  38 & --- & --- &  46 & --- \\
Ni\,{\sc i} & 6204.61 & 4.09 & $-$1.150 & MFK & --- &  34 & --- &  49 &  56 \\
Ni\,{\sc i} & 6223.99 & 4.11 & $-$0.980 & MFK & --- & --- & --- &  50 & --- \\
Ni\,{\sc i} & 6230.10 & 4.11 & $-$1.260 & MFK & --- &  27 & --- & --- & --- \\
Ni\,{\sc i} & 6327.60 & 1.68 & $-$3.150 & MFK &  60 &  74 & --- &  93 &  99 \\
Ni\,{\sc i} & 6482.81 & 1.94 & $-$2.630 & MFK & --- &  82 & 112 &  96 & --- \\
Ni\,{\sc i} & 6586.32 & 1.95 & $-$2.810 & MFK &  76 & --- &  92 &  76 & --- \\
Ni\,{\sc i} & 6635.14 & 4.42 & $-$0.830 & MFK & --- & --- & --- &  31 &  43 \\
Ni\,{\sc i} & 6643.64 & 1.68 & $-$2.030 & MFK & --- & 129 & --- & 139 & --- \\
Ni\,{\sc i} & 6767.78 & 1.83 & $-$2.170 & MFK & 131 & --- & --- & --- & 155 \\
\tableline
\end{tabular}
\end{table}

\begin{table*}[!h]
\noindent
\tiny
\begin{tabular}{cccccccccc}
\multicolumn{10}{c}{Table 8, continued.}\\
\tableline\tableline
    & & & & &\multicolumn{5}{c}{Equivalent Widths (m\AA)} \\\tableline
\multicolumn{5}{c}{} & \multicolumn{5}{c}{Star}\\
\cline{6-10}
Element & $\lambda$ & $\chi$(eV) & $\log gf$ & Ref & 1534 & 1616 & 1349 & 784 & 663 \\
\tableline
Ni\,{\sc i} & 6772.32 & 3.66 & $-$0.970 & R03 & --- & --- & --- &  91 &  90 \\
Ni\,{\sc i} & 7788.93 & 1.95 & $-$1.990 & E93 & --- & 118 & 156 & 142 & --- \\
\tableline
\\
\end{tabular}
\par References: B86: Blackwell D.E. et al. (1986); Ca07: Carretta et al. (2007);
\par D2002; Depagne et al. (2002); DS91: Drake \& Smith (1991);
\par E93: Edvardsson et al. (1993); GS: Gratton \& Sneden (1988);
\par MFK: Martin et al., (2002); MR94: Mcwilliam \& Rich (1994);
\par PS: Preston \& Sneden (2001); R03: Reddy et al. (2003);
\par R99: Reddy et al. (1999); WSM: Wiese, Smith \& Miles (1969).

\end{table*}

\clearpage

\begin{landscape}
\begin{table*}
\footnotesize
\caption{ Atmospheric parameters from photometry (ph) and spectroscopy (sp).}
\centering
\label{tab:atmparam}
\begin{tabular}{ccccccccc}\\\tableline\tableline
ID     &  $T_{\rm eff, ph}$ & log~$g_{ph}$  &  $T_{\rm eff, sp}$ & log~$g_{sp}$  & $\xi$         & $[FeI/H]\pm$ $\sigma$ (\#)              & $[FeII/H]\pm$ $\sigma$ (\#)   & Comment    \\
       &    (K)             & dex           &    (K)             & dex           & km\,s$^{-1}$  &                                         &                               &            \\\tableline 
395    &      5205          &   2.74        &     5100           &  2.7          &   1.6         &    $-$0.15$\pm$0.15(53)                 &    $-$0.15$\pm$0.12(5)        &   Member   \\
663    &      5123          &   2.85        &     4900           &  3.2          &   2.3         &       0.07$\pm$0.14(37)                 &       0.06$\pm$0.13(3)        &   Member   \\
769    &      5196          &   2.53        &     5200           &  3.0          &   1.3         &       0.10$\pm$0.14(53)                 &       0.08$\pm$0.17(3)        &   Member   \\
784    &      4955          &   2.82        &     5000           &  2.5          &   1.7         &    $-$0.08$\pm$0.13(39)                 &    $-$0.11$\pm$0.12(6)        &  Non-Member\\
806    &      6324          &   2.86        &     6000           &  2.7          &   4.9         &    $-$0.53$\pm$0.12(20)                 &    $-$0.52(2)                 &  Non-Member/binary Cepheid?\\
1110   &      5161          &   2.68        &     5350           &  3.4          &   1.0         &       0.03$\pm$0.15(44)                 &       0.01$\pm$0.09(6)        &   Member   \\
1118   &      4958          &   2.63        &     5100           &  2.6          &   1.5         &    $-$0.16$\pm$0.12(44)                 &    $-$0.17$\pm$0.10(3)        &   Member   \\
1349   &      4796          &   2.72        &     5100           &  2.6          &   2.1         &       0.01$\pm$0.19(25)                 &       0.01(2)                 &   Member   \\
1534   &      4982          &   2.73        &     5000           &  2.0          &   2.2         &    $-$0.30$\pm$0.16(15)                 &    $-$0.29(2)                 &  Non-Member\\
1616   &      5465          &   2.83        &     5450           &  3.5          &   1.7         &    $-$0.07$\pm$0.16(31)                 &    $-$0.09$\pm$0.14(5)        &  Non-Member\\\tableline
\\
\end{tabular}

\par \textbf{Notes.} For [Fe I/H] and [Fe II/H], we also show the standard deviation and the number of lines employed.

\end{table*}
\end{landscape}

\clearpage

\begin{table}
\caption{Adopted solar abundances.}
\tabcolsep 0.33truecm
\label{sun}
\begin{tabular}{lccc}\\\tableline\tableline
Element              & This  & Grevesse \&  & Asplund                \\
$_{\rule{0pt}{8pt}}$ & work  & Sauval (1998)& et al. (2009) \\
\tableline     
Fe                   & 7.50  & 7.50         & 7.50                   \\ 
Na                   & 6.26  & 6.33         & 6.24                   \\
Mg                   & 7.55  & 7.58         & 7.60                   \\
Al                   & 6.31  & 6.47         & 6.45                   \\
Si                   & 7.61  & 7.55         & 7.51                   \\
Ca                   & 6.37  & 6.36         & 6.34                   \\
Ti                   & 4.93  & 5.02         & 4.95                   \\
Cr                   & 5.65  & 5.67         & 5.64                   \\
Ni                   & 6.29  & 6.25         & 6.22                   \\
Y                    & 2.04  & 2.24         & 2.21                   \\
Ba                   & 2.18  & 2.13         & 2.18                   \\
Ce                   & 1.48  & 1.58         & 1.58                   \\
Nd                   & 1.42  & 1.50         & 1.42                   \\
\tableline\\
\end{tabular}

\end{table}

\clearpage

\begin{landscape}
\begin{table*}
\footnotesize
\caption{Abundance Ratios ($[X/Fe]$) for the elements from Na to Cr for the stars observed.}
\centering
\label{abunda-Na}
\begin{tabular}{lccccccc}\\
\tableline\tableline
\multicolumn{8}{c}{Cluster giants}\\\tableline
ID            & [Na/Fe]NLTE       & [Mg/Fe]            & [Al/Fe]            & [Si/Fe]             & [Ca/Fe]             & [Ti/Fe]             & [Cr/Fe]            \\\tableline
395           & +0.15(2)          &$+$0.10$\pm$0.13(3) &     ---            &$+$0.13$\pm$0.09(4)  &$+$0.02$\pm$0.13(7)  &$+$0.07(1)           &$+$0.02$\pm$0.14(7)  \\
663           & +0.12(2)          &$+$0.03$\pm$0.12(3) &$+$0.15$\pm$0.04(3) &$+$0.23$\pm$0.11(4)  &$-$0.06(2)           &$+$0.04$\pm$0.14(9)  &$-$0.10$\pm$0.02(3)  \\
769           & +0.04(2)          &$-$0.12$\pm$0.13(4) &     ---            &$-$0.05$\pm$0.04(5)  &$-$0.05$\pm$0.15(7)  &$+$0.10$\pm$0.11(18) &$+$0.04$\pm$0.12(7)  \\
1110          & +0.13(2)          &$+$0.03$\pm$0.14(3) &     ---            &$-$0.14(2)           &$+$0.04$\pm$0.11(5)  &$+$0.09$\pm$0.13(11) &$+$0.18$\pm$0.06(6)  \\
1118          & +0.38(2)          &$+$0.23$\pm$0.07(4) &     ---            &$+$0.07$\pm$0.09(3)  &$+$0.12$\pm$0.08(5)  &$+$0.08$\pm$0.13(13) &$+$0.14$\pm$0.13(6)  \\
1349          & ---               &$-$0.11(2)          &$+$0.16(2)          &$-$0.19(2)           &$-$0.01(2)           &$+$0.28$\pm$0.11(4)  &$+$0.30$\pm$0.09(5)  \\\tableline
Tombaugh 1$^a$& +0.17$\pm$0.06    &$+$0.03$\pm$0.05    &$+$0.15             &$+$0.01$\pm$0.07     &$+$0.01$\pm$0.03     &$+$0.11$\pm$0.04     &$+$0.10$\pm$0.06     \\\tableline\tableline
\multicolumn{8}{c}{Field stars}\\\tableline
ID            & [Na/Fe]NLTE       & [Mg/Fe]            & [Al/Fe]            & [Si/Fe]             & [Ca/Fe]             & [Ti/Fe]             & [Cr/Fe]            \\\tableline
784           & +0.20(2)          &$+$0.17$\pm$0.13(4) &$+$0.27(2)          &$+$0.14$\pm$0.10(5)  &$-$0.06$\pm$0.09(6)  &$+$0.03$\pm$0.07(10) &$-$0.13$\pm$0.10(5)  \\
806$^b$       & +0.27(1)          &$+$0.02(1)          &     ---            &$+$0.14$\pm$0.11(4)  &$+$0.05(2)           &$+$0.32(1)           &$-$0.12$\pm$0.08(3)  \\
1534          & +0.21(1)          &$+$0.25$\pm$0.02(3) &     ---            &$-$0.15$\pm$0.09(5)  &$-$0.20$\pm$0.15(4)  &$+$0.01(2)           &$-$0.22$\pm$0.11(3)  \\
1616          & +0.34(1)          &$+$0.18$\pm$0.03(3) &$-$0.10(2)          &$-$0.21$\pm$0.08(3)  &$-$0.01$\pm$0.06(3)  &$+$0.27$\pm$0.13(10) &$+$0.17$\pm$0.14(4)  \\\tableline
\tableline\\
\end{tabular}
\flushleft
\par \textbf{Notes.} For all abundances ratios, we also show the standard deviation and the number of lines employed. [Na/Fe] accounts for the NLTE effects calculated as in Gratton et al. (1999), see text. 
(a) Mean abundance ratio for each element for Tombaugh 1. (b) The classical Cepheid XZ CMa.
\end{table*}
\end{landscape}

\clearpage

\begin{table*} 
\caption{Abundance Ratios ($[X/Fe]$) for the elements from Ni to Nd for the stars observed.}
\centering
\label{abunda-Ni}
\begin{tabular}{lccccc}
\\\tableline\tableline
\multicolumn{6}{c}{Cluster giants}\\\tableline
ID            &  [Ni/Fe]           & [Y/Fe]             & [Ba/Fe]             & [Ce/Fe]             & [Nd/Fe]             \\\tableline 
395           &$-$0.06$\pm$0.13(18)&$-$0.10(1)          &$+$0.28(1)           &$+$0.25(1)           &$+$0.35(1)           \\
663           &$-$0.04$\pm$0.15(14)&$+$0.00(1)          &      ---            &      ---            &      ---            \\
769           &$+$0.02$\pm$0.15(28)&$+$0.07(1)          &$+$0.38(1)           &$+$0.10(1)           &$+$0.30(1)           \\
1110          &$-$0.05$\pm$0.13(19)&$+$0.14(1)          &$+$0.37(1)           &$+$0.37(1)           &$+$0.47(1)           \\
1118          &$+$0.04$\pm$0.12(19)&$+$0.13(1)          &$+$0.36(1)           &$+$0.29(1)           &      ---            \\
1349          &$-$0.14$\pm$0.09(12)&      ---           &      ---            &      ---            &      ---            \\\tableline
Tombaugh 1$^a$&$-$0.04$\pm$0.02    &$+$0.06$\pm$0.04    &$+$0.35$\pm$0.03     &$+$0.25$\pm$0.06     &$+$0.37$\pm$0.05     \\\tableline\tableline
\multicolumn{6}{c}{Field stars}\\\tableline
ID            &  [Ni/Fe]           & [Y/Fe]             & [Ba/Fe]             & [Ce/Fe]             & [Nd/Fe]            \\\tableline 
784           &$-$0.14$\pm$0.14(18)&$+$0.30(1)          &$-$0.02(1)           &$+$0.20(1)           &$+$0.32(1)           \\
806$^b$       &$+$0.20$\pm$0.16(3) &      ---           &$+$0.73(1)           &      ---            &      ---            \\
1534          &$-$0.27$\pm$0.13(9) &$-$0.07(1)          &$-$0.05(1)           &      ---            &      ---            \\
1616          &$+$0.02$\pm$0.13(15)&$-$0.11(1)          &$+$0.39(1)           &      ---            &$+$0.04(1)           \\\tableline
\tableline\\
\end{tabular}
\flushleft
\par \textbf{Notes.} For all abundances ratios, we also show the standard deviation and the number of lines employed. 
(a) Mean abundance ratio for each element for Tombaugh 1. (b) The classical Cepheid XZ CMa.
\end{table*}

\clearpage

\begin{table}
\caption{Abundance uncertainties for star 769.} 
\tabcolsep 0.33truecm
\label{error}
\begin{tabular}{lcccc}\\\tableline\tableline
Element & $\Delta T_{eff}$ & $\Delta\log g$ & $\Delta\xi$ & $\left( \sum \sigma^2 \right)^{1/2}$ \\
$_{\rule{0pt}{8pt}}$ & $+$180~K & $+$0.3 & $+$0.3 km\,s$^{-1}$ &  \\
\tableline     
Fe\,{\sc i}    & $+$0.14  &    0.00 & $-$0.14 & 0.20 \\ 
Fe\,{\sc ii}   & $-$0.11  & $+$0.19 & $-$0.11 & 0.25 \\
Na\,{\sc i}    & $+$0.12  & $-$0.01 & $-$0.05 & 0.13 \\
Mg\,{\sc i}    & $+$0.09  & $-$0.01 & $-$0.07 & 0.11 \\
Al\,{\sc i}    & $+$0.09  & $-$0.04 & $-$0.05 & 0.11 \\
Si\,{\sc i}    &    0.00  & $+$0.05 & $-$0.05 & 0.07 \\
Ca\,{\sc i}    & $+$0.16  & $-$0.04 & $-$0.15 & 0.22 \\
Ti\,{\sc i}    & $+$0.23  & $-$0.02 & $-$0.12 & 0.26 \\
Cr\,{\sc i}    & $+$0.15  & $-$0.02 & $-$0.09 & 0.18 \\
Ni\,{\sc i}    & $+$0.11  & $+$0.03 & $-$0.12 & 0.17 \\
Y\,{\sc ii}    & $-$0.02  & $+$0.08 & $-$0.05 & 0.10 \\
Zr\,{\sc i}    & $+$0.01  & $-$0.03 & $-$0.06 & 0.07 \\
Ba\,{\sc ii}   & $+$0.12  & $+$0.17 & $-$0.14 & 0.25 \\
Ce\,{\sc ii}   & $-$0.02  & $+$0.05 & $-$0.10 & 0.11 \\
Nd\,{\sc ii}   & $+$0.05  & $+$0.15 & $-$0.05 & 0.17 \\
\tableline\\
\end{tabular}

\par \textbf{Notes.} Each column gives the variation of the abundance caused by the variation in $T_{\rm eff}$, $\log g$ and $\xi$. The last column gives the compounded rms uncertainty of the second to fourth columns. Abundance uncertainties of aluminium were calculated using the star 663.

\end{table}

\end{document}